\documentclass[%
twocolumn,
superscriptaddress,
amsmath,
amssymb, 
aps,
prl,
floatfix,
]{revtex4-2}

\usepackage{array}
\usepackage{graphicx}
\usepackage{bm}
\usepackage{amsmath}
\usepackage[dvipsnames]{xcolor}
\usepackage{hyperref}
\graphicspath{{./figures/}}

\begin{document}

\title{Tunable Three-Body Interactions in Microwave-Shielded Polar Molecules}
\author{Hanwei Yang}
\affiliation{Institute for Molecules and Materials, Radboud University, Nijmegen, The Netherlands}
\author{Tijs Karman}
\email{t.karman@science.ru.nl}
\affiliation{Institute for Molecules and Materials, Radboud University, Nijmegen, The Netherlands}

\begin{abstract}
We show that microwave shielding, which enables collisionally stable ultracold polar molecules,
also leads to genuine nonadditive three-body interactions comparable to the shielding interaction itself.
Their magnitude, sign, and geometrical dependence can be flexibly controlled through microwave parameters and orientation. 
Interaction strengths can reach the ten kilohertz-scale for ultrapolar molecules in typical optical lattices.
The nonadditive three-body interactions can stabilize novel many-body phases,
and should be accounted for in quantum simulation with microwave shielded molecules.
\end{abstract}

\maketitle

Ultracold polar molecules provide a versatile platform for exploring quantum many-body physics owing to their long-range, anisotropic interactions and rich internal level structure \cite{gadway:2016,bohn:2017,cornish:2024}. Progress in direct cooling \cite{langen:2024} and molecular assembly from ultracold atoms \cite{ni:2008,park:2015,zhang:2021} has brought molecular gases into the quantum-degenerate regime \cite{demarco:2019}. 
Microwave shielding \cite{karman:2018} has greatly suppressed collisional losses~\cite{bigagli:2023,lin:2023,yuan2025extreme}, allowing efficient evaporative cooling to quantum degeneracy\cite{schindewolf:2022,shi:2025}, while providing flexible control over long-range dipolar interactions~\cite{deng:2023,karman:2025}.
These advances have enabled Bose--Einstein condensation \cite{bigagli:2024,shi:2025} and the formation of strongly interacting self-bound droplets and droplet arrays \cite{zhang:2026}.
Other shielding schemes have also been proposed \cite{xie.T,karam,augustovicova},
including static field shielding~\cite{gonzalez} that has similarly been realized to suppress collisional loss, enable evaporative cooling, and reach degeneracy \cite{matsuda2020resonant,ye:2026,bakr:2026}.
Combining these capabilities with blockade-assisted loading schemes further opens the way to preparing optical lattices and tweezer arrays at high filling and low entropy \cite{karman:2026blockade}, providing a credible route to the simulation of extended Hubbard models \cite{Dutta_2015,Chanda_2025}.

Together with recent advances in molecular quantum gas microscopy that enable site-resolved measurements \cite{christakis:2023,mortlock:2026},
shielding offers new opportunities to observe and simulate quantum many-body physics with ultracold molecules.
In optical lattices, parameters were recently studied for extended Hubbard models with microwave-tunable on-site and off-site interactions \cite{li:2026hubbard,perez:2026hubbard,stewartwiese:2026hubbard}.
In bulk gases, three-body recombination into field-linked bound states \cite{stevenson:2024,xinyu+shi} and Efimov physics with microwave-shielded molecules have been investigated \cite{singh:2026}.
In all cases, the descriptions are based on effective~\cite{deng:2023} two-body interactions.

However, substantial nonadditive many-body interactions can arise for microwave-shielded molecules~\cite{prieschl:2026}.
Double microwave shielding~\cite{karman:2025} naturally realizes a scenario similar to that proposed by B{\"u}chler \textit{et al.}~\cite{buechler:2007}, who showed that a combination of microwave and static electric fields can generate strong three-body interactions and proposed Hubbard models in which such terms become dominant.

\begin{figure*}
    \centering
    \includegraphics[width=0.95\linewidth]{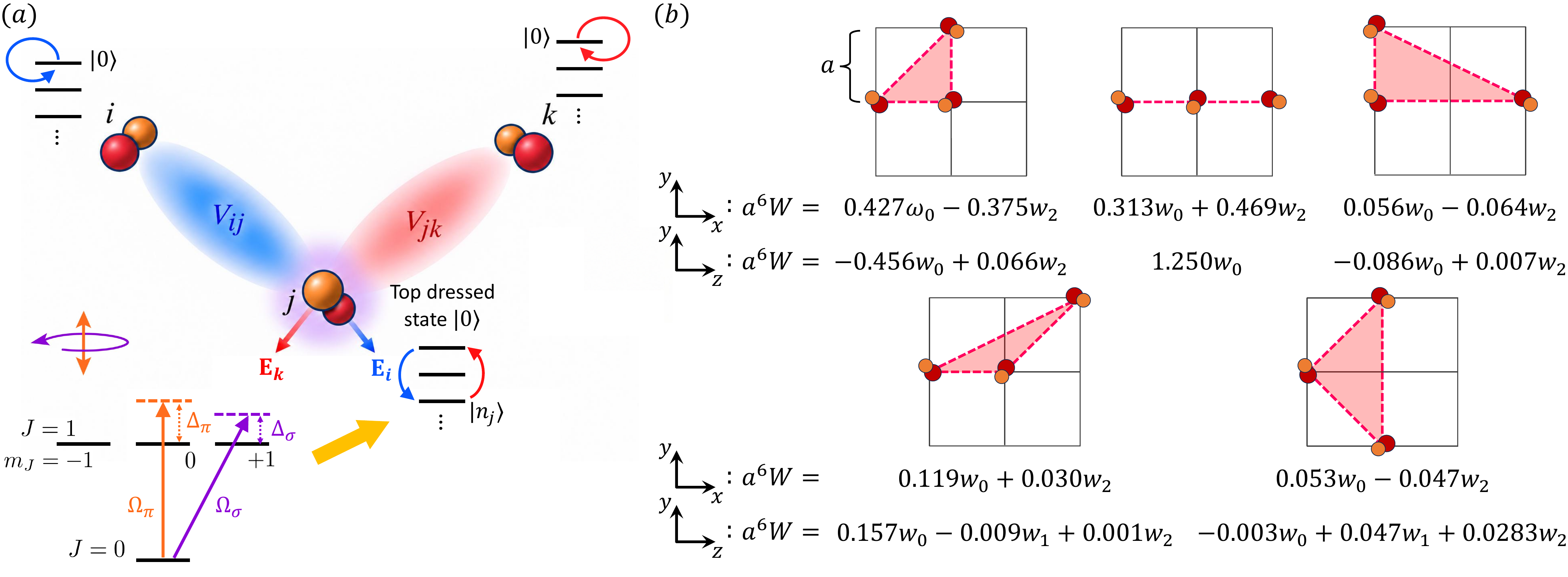}
     \caption{{\bf Three-body interactions in microwave-shielded polar molecules}. (a) Schematic of the mechanism giving rise to the three-body interaction. Microwave dressing mixes rotational states of opposite parity, producing a set of dressed states with tunable intermolecular interactions. The dipolar fields generated by molecules $i$ and $k$ act simultaneously on the molecule $j$, and induce a virtual transition of molecule $j$ to an intermediate state $|n_j\rangle$ by $\hat V_{ij}$ and back through $\hat V_{jk}$, and vice versa, while $i$ and $k$ themselves remain in the initial state. 
     (b) Representative three-molecule configurations of a two-dimensional square lattice in the $xy$ or $zy$ planes, with the microwave quantization axis fixed along $\hat z$. The red dashed lines indicate pairwise two-body interactions, while the shaded areas represent the three-body interactions. 
     }
    \label{fig1}
\end{figure*}

In this Letter, we show that microwave shielding naturally realizes genuine nonadditive three-body interactions arising from parity mixing induced by the dressing fields.
We derive the general form of the resulting effective potential.
By tuning the microwave parameters, three-body interactions can become comparable to pairwise contributions and even dominate when the latter are suppressed, without compromising collisional shielding \cite{karman:2018,karman:2025}.
The three-body interaction can also be strong in an absolute sense, reaching the ten kilohertz scale that is convenient for quantum gas microscopy in optical lattices,
especially for ultrapolar molecules such as KAg~\cite{smialkowski:2021}. Compared with the scheme proposed by B{\"u}chler \textit{et al.} \cite{buechler:2007}, double microwave shielding offers greater flexibility in tuning the strength of the three-body interaction relative to pairwise terms.
Furthermore, the geometrical dependence also becomes controllable through microwave parameters and field orientation. 
These results establish microwave-shielded dipolar molecules as a promising platform for exploring controllable nonadditive three-body interactions in few~\cite{stevenson:2024,singh:2026} and many-body physics \cite{bonnes:2010,bonnes:2011b,mazza:2010,singh:2016}.

To illustrate the origin of the nonadditive three-body interactions,
consider molecule $j$ subject to the dipolar fields $\mathbf{E}_i$
and $\mathbf{E}_k$ generated by molecules $i$ and $k$, as shown
in Fig.~\ref{fig1}(a). In a simple classical picture, the induced
energy shift of polarizable molecule $j$ is proportional to
$|\mathbf{E}_i+\mathbf{E}_k|^2$.
Direct terms proportional to $|\mathbf{E}_i|^2,~|\mathbf{E}_k|^2$ contribute pairwise interactions, whereas the cross term $\mathbf{E}_i\cdot\mathbf{E}_k$ depends on all three molecules and results in a genuine three-body interaction.

The classical cross term above has a direct quantum-mechanical counterpart in second-order perturbation theory. We consider all the molecules initially prepared in the top dressed state $|0\rangle$ under microwave dressing, thereby enabling collisional shielding~\cite{karman:2018,karman:2025}. The overall interactions among three molecules are $\hat{V}_{ij}+\hat{V}_{jk}+\hat{V}_{ki}$, where $\hat{V}_{ij}$ denotes the dipole-dipole interaction between molecules $i$ and $j$. For the incoming state $|000\rangle$, the second-order energy shift is
\begin{equation}
\sum_{n_i n_j n_k}
\frac{
\big |\langle 000|\hat V_{ij}+\hat V_{jk}+\hat V_{ki}|n_i n_j n_k\rangle \big |^2
}{
E_{000}-E_{n_i n_j n_k}
}.
\label{2nd}
\end{equation}
Here $|n_i n_j n_k\rangle$ runs over all three-molecule dressed states other than $|000\rangle$ with the asymptotic energy $E_{n_i n_j n_k}$. 
Terms such as $\hat V_{ij}\hat V_{ij}$, which involve the same molecular pair twice, yield the usual second-order pairwise interactions.
Cross terms involving distinct pairs, such as $\hat V_{ij} \hat V_{jk}$,
contribute to the nonadditive part of the energy shift~\cite{buechler:2007}.

We take the cross terms involving $\hat V_{ij}$ and $\hat V_{jk}$ as an example. 
Since only $\hat V_{ij}$ acts on molecule $i$,
and only $\hat V_{jk}$ acts on molecule $k$,
these molecules cannot undergo a virtual excitation.
Only molecule $j$ occurs in both interactions and can be virtually excited and de-excited, and hence contribute to the three-body interaction
\begin{equation}
\begin{aligned}
W_{ij,jk}^{(2)}
={}&
\sum_{|n_j\rangle\neq |0\rangle}
\Big\{\frac{
\langle000|\hat V_{ij}|0n_j0\rangle
\langle0n_j0|\hat V_{jk}|000\rangle
}{
E_{000}-E_{0n_j0}
}
\\
&+
\frac{
\langle000|\hat V_{jk}|0n_j0\rangle
\langle0n_j0|\hat V_{ij}|000\rangle
}{
E_{000}-E_{0n_j0}
}\Big\}.
\label{cross}
\end{aligned}
\end{equation}
The resulting interaction depends simultaneously on the two intermolecular vectors $\hat{R}_{ij}$,~$\hat{R}_{jk}$ and scales as $R_{ij}^{-3}R_{jk}^{-3}$, reflecting its genuine three-body character. Other analogous contributions are obtained by taking $i$ or $k$ as the virtually excited molecule, corresponding to $\hat V_{ij}\hat V_{ki},~\hat V_{ki}\hat V_{ij}$ and $\hat V_{jk}\hat V_{ki},~\hat V_{ki}\hat V_{jk}$ in Eq.~\eqref{2nd}.

\begin{figure*}
    \centering
    \includegraphics[width=1.0\linewidth]{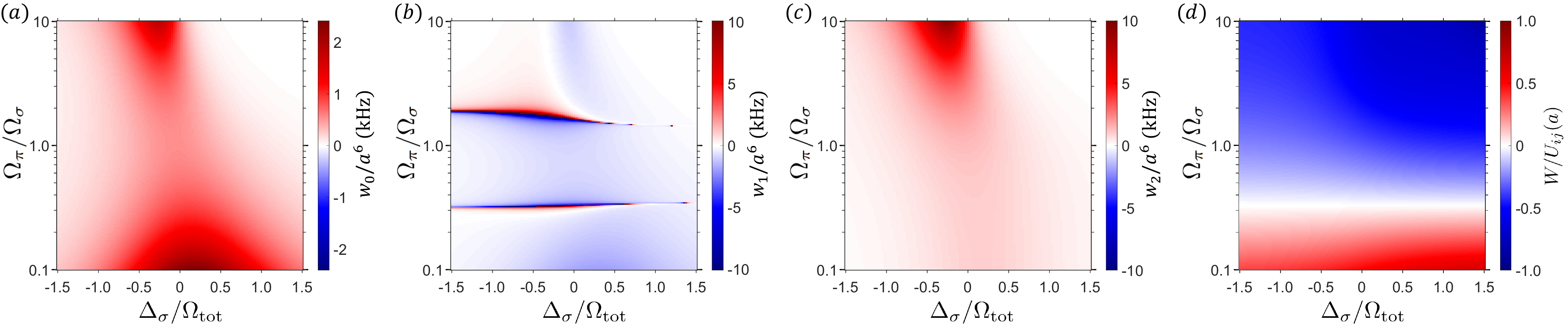}
     \caption{{\bf Tunability of the three-body interaction} for KAg molecules under double microwave shielding. {(a)–(c)} Three-body coefficients $w_0/a^6$, $w_1/a^6$, and $w_2/a^6$, respectively, as functions of the Rabi-frequency ratio $\Omega_\pi/\Omega_\sigma$ and normalized detuning $\Delta_\sigma/\Omega_{\rm tot}$, with the lattice constant $a=532\,\mathrm{nm}$ and fixed total Rabi frequency $\Omega_{\rm tot}=1\,\mathrm{MHz}$, while $\Delta_\pi$ is determined by compensation of the first-order dipolar interaction. {(d)} The ratio of the total three-body interaction $W$ to the remaining pairwise interaction $U_{ij}$ between two molecules separated by $a$, evaluated for the first configuration in Fig.~\ref{fig1}(b) in the $xy$ plane.}
    \label{fig2}
\end{figure*}

As illustrated in Fig.~\ref{fig1}(a), the microwave dressing mixes rotational states of opposite parity~\cite{karman:2018,karman:2025}, making the diagonal dipole moments of initial state $|0\rangle$ nonzero, and thereby enabling the three-body interaction in Eq.~\eqref{cross}. 
For comparable separations $R_{ij}\sim R_{jk}\sim R_{ki}\sim R$, this contribution scales as $R^{-6}$, like the second-order pairwise interaction. 
At the compensation point of double microwave shielding~\cite{karman:2025},
the first-order dipolar interaction vanishes,
which is relevant for eliminating three-body recombination~\cite{stevenson:2024} and for obtaining stable quantum gases~\cite{bigagli:2024,zhang:2026}.
Under this condition, the three-body interaction is of the same order as the remaining pairwise potential.
Without microwave dressing, the molecular states possess definite parity and the three-body interaction vanishes to second order.
The leading nonadditive three-body contribution then arises at third order as an Axilrod--Teller--Muto (ATM) type interaction~\cite{axilrod:1943,muto:1943},
associated with the cyclic product $\hat V_{ij}\hat V_{jk}\hat V_{ki}$ that flips the parity of each molecule twice and decays much more rapidly as $R^{-9}$.

The angular dependence of Eq.~\eqref{cross} follows from the rank-2 tensor structure of the dipole-dipole interaction \cite{micheli:2007,karman:2025}. Collecting the independent angular contributions with $|m|=0,1,2$ into coefficients $w_0$, $w_1$ and $w_2$, and summing the contributions corresponding to three choices of the virtually excited molecule, the total effective three-body interaction can be written as
\begin{align}
W&=
\big\{
w_0 C_{20}(\hat{ R}_{ij})C_{20}(\hat{R}_{jk})
+w_1\,\mathrm{Re}[
C_{21}(\hat{R}_{ij})
C_{2,-1}(\hat{R}_{jk})
]
\nonumber\\
&
+w_2\,\mathrm{Re}[
C_{22}(\hat{R}_{ij})
C_{2,-2}(\hat{R}_{jk})
]
\big\}/R_{ij}^{3}R_{jk}^{3}
+\mathrm{perm.},
\label{three}
\end{align}
where $C_{2m}$ are Racah-normalized spherical harmonics, with the polar angles defined relative to the microwave quantization axis. The coefficients $w_0$, $w_1$, and $w_2$ contain the dependence on microwave dressing, while the spherical harmonics encode the spatial configuration of the three molecules, allowing the microwave and geometrical dependences to be treated separately.
We provide Python code to evaluate both the microwave-dependent coefficients and geometrical factors, and compute the resulting three-body potential for arbitrary microwave parameters and molecular configurations (see Supplemental Material)~\cite{gitlab}.

To illustrate the dependence of the three-body interaction on both molecular geometry and microwave dressing, we consider molecules confined to a two-dimensional square lattice with lattice constant $a$. For any fixed three-molecule configuration, the total three-body interaction can be written as
$W=\left(\gamma_0 w_0+\gamma_1 w_1+\gamma_2 w_2\right)/a^6$,
where the dimensionless geometrical factors \(\gamma_0\), \(\gamma_1\) and \(\gamma_2\) are independent of the microwave parameters.
Fig.~\ref{fig1}(b) shows several representative compact three-molecule configurations for which the three-body interaction is appreciable. 
We compare these configurations with the square lattice in either the $xy$ or $zy$ plane, with the microwave quantization axis fixed along $\hat{z}$.
These results show that the three-body interaction depends strongly on both the molecular configuration and its orientation relative to the microwave quantization axis. In particular, when the quantization axis is perpendicular to the lattice plane, the contribution proportional to $w_1$ vanishes for any three-molecule configuration.

We now consider how the three-body interaction can be tuned through the microwave parameters. The two microwave fields provide independently tunable Rabi frequencies $\Omega_\sigma,\Omega_\pi$ and detunings $\Delta_\sigma,\Delta_\pi$. To illustrate the resulting tunability, Figure~\ref{fig2}(a)-(c) shows the coefficients $w_0/a^6$, $w_1/a^6$ and $w_2/a^6$ for KAg molecules with $a=532\,\mathrm{nm}$ and fixed total Rabi frequency $\Omega_{\rm tot}=(\Omega_\pi^2+\Omega_\sigma^2)^{1/2}=1\,\mathrm{MHz}$. We vary $\Omega_\pi/\Omega_\sigma$ and $\Delta_\sigma/\Omega_{\rm tot}$ as the independent variables, while adjusting $\Delta_\pi$ at each point to cancel the first-order dipolar interaction. Here $w_0$ and $w_2$ remain positive, whereas $w_1$ is predominantly negative and exhibits narrow resonant features associated with near-degeneracies with other Floquet-dressed states \cite{karman:2025}. 
These variations allow the overall three-body interaction to be tuned over a broad range, from nearly zero to the ten kHz scale, combining sizable interaction strengths with broad and flexible control through the microwave parameters.

Although Fig.~\ref{fig2}(a)-(c) shows results for KAg, the dependence on the scaled microwave parameters is universal~\cite{Dutta_PRR,jozwiakmap2026}, with the three-body coefficients taking the form
\begin{equation}
w_m=\frac{d^4}{\Omega_{\rm tot}}
f_m\!\left(
\frac{\Omega_\pi}{\Omega_\sigma},
\frac{\Delta_\sigma}{\Omega_{\rm tot}},
\frac{\Delta_\pi}{\Omega_{\rm tot}}
\right),\quad m=0,1,2,
\label{wi}
\end{equation}
where $d$ is the body-fixed dipole moment and the dimensionless functions $f_m$ depend only on the scaled microwave parameters, of which only two are independent at compensation. 
The same interactions apply to different molecular species and at different total Rabi frequencies,
apart from an overall scaling with $d^4/\Omega_{\rm tot}$.

\begin{figure}[b]
    \centering
    \includegraphics[width=1.0\linewidth]{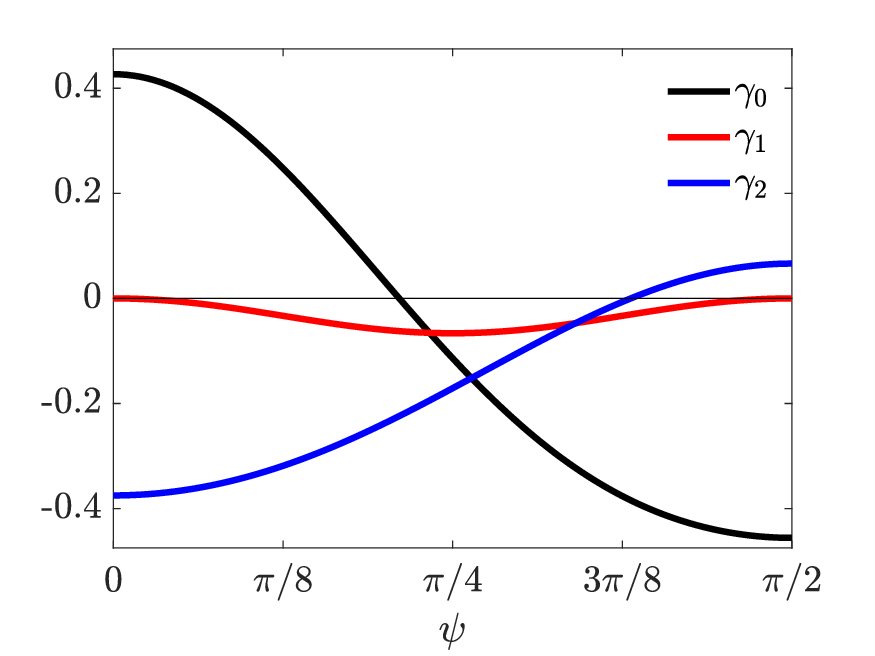}
    \caption{{\bf Orientation dependence of the three-body interaction} for the first configuration shown in Fig.~\ref{fig1}(b). The microwave quantization axis is fixed along $\hat{z}$. 
    The lattice is rotated counter-clockwise about the $y$ axis over an angle $\psi.$
    At $\psi=0$ ($\pi/2$), the lattice is in the $xy$ ($zy$) plane, repsectively.}
    \label{fig3}
\end{figure}

Figure~\ref{fig2}(d) further compares the total three-body interaction $W$ with the pair interaction $U_{ij}(a)$~\cite{yang:2026polaron} between two nearest-neighbor molecules, for the first configuration in Fig.~\ref{fig1}(b) with the lattice in the $xy$ plane. At the compensation point, where the first-order dipolar interaction vanishes, the three-body contribution reaches a substantial fraction of the pairwise interaction for a wide variety of microwave settings, although it remains smaller in magnitude throughout the parameter range shown. While the second-order pairwise interaction is repulsive, the anisotropic three-body contribution can be either attractive or repulsive. For the top dressed incoming channel, however, Eq.~\eqref{2nd} ensures that the total second-order interaction, including all pairwise and three-body contributions, remains positive for any molecular configuration. Thus, an attractive three-body contribution cannot overcome the combined second-order pairwise repulsion, consistent with the preservation of collisional shielding. Nevertheless, the three-body interaction can still substantially affect few-body energetics. For example, the first two configurations in Fig.~\ref{fig1}(b) have nearly the same pairwise interaction but markedly different three-body contributions, highlighting their strong dependence on molecular geometry.

\begin{figure}
    \centering
    \includegraphics[width=1.0\linewidth]{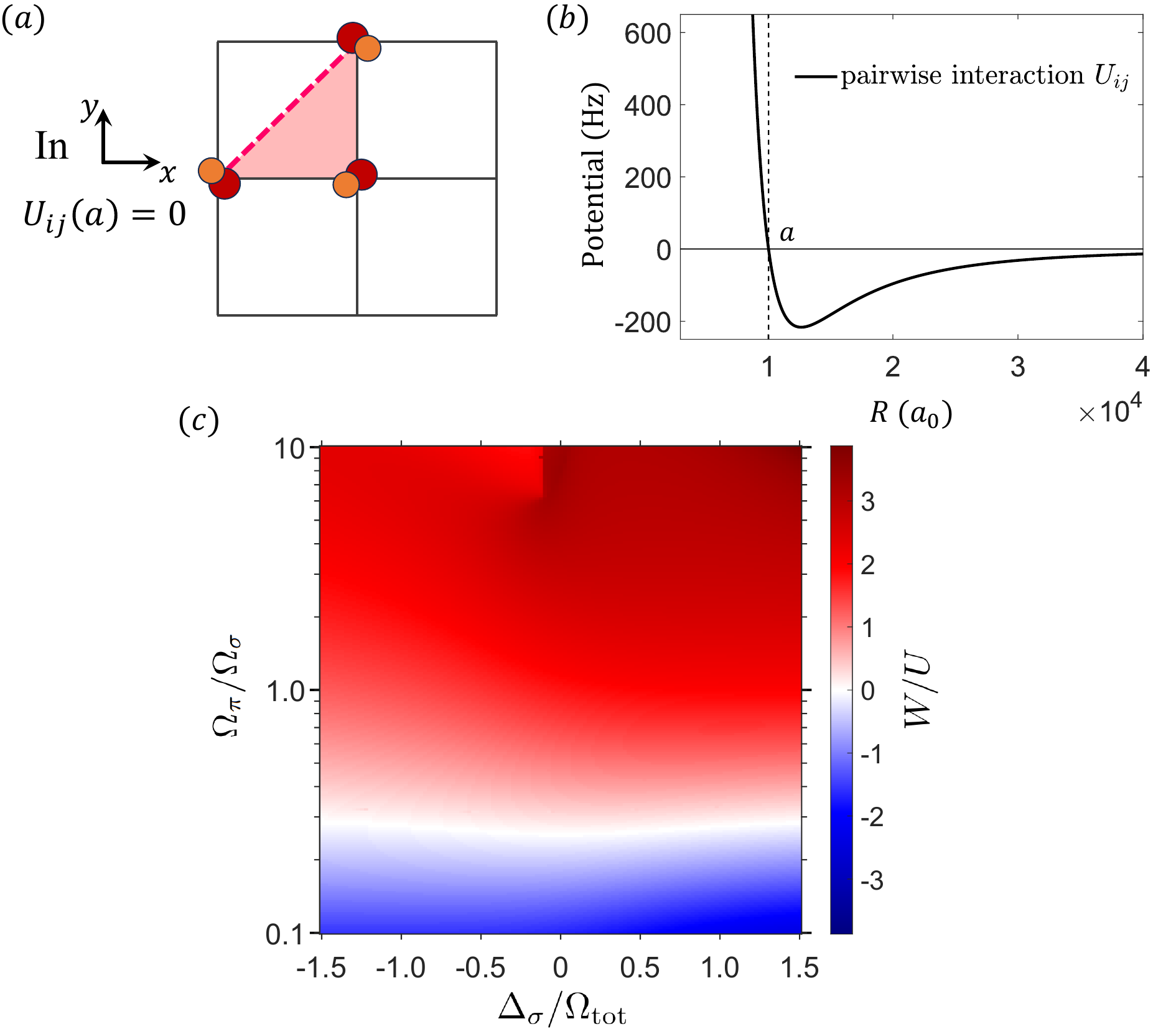}
    \caption{{\bf Three-body interactions for canceled nearest-neighbor interactions}. (a) For the configuration shown only the next-nearest-neighbor pair at separation $\sqrt{2}a$ contributes to the residual pairwise interaction, while the three-body interaction remains sizable. 
    (b) Schematic effective pairwise potential $U_{ij}(R)$ as a function of intermolecular separation $R$, with a zero crossing at $R=a$.
    (c) For KAg molecules, the ratio of the total three-body interaction $W$ to the residual pairwise interaction $U$ for the configuration in (a), mapped over $\Omega_\pi/\Omega_\sigma$ and $\Delta_\sigma/\Omega_{\rm tot}$, with $a=532\,\mathrm{nm}$ and $\Omega_{\rm tot}=1\,\mathrm{MHz}$. For each set of these parameters, $\Delta_\pi$ is adjusted to enforce $U_{ij}(a)=0$.
}
    \label{fig4}
\end{figure}

Next, we investigate the effect of relative orientation of the lattice and microwave polarization.
We fix the microwave quantization axis along $\hat{z}$ and continuously rotate the lattice plane counterclockwise about $\hat{y}$ by an angle $\psi$, from the $xy$ plane at $\psi=0$ to the $zy$ plane at $\psi=\pi/2$. 
Figure~\ref{fig3} shows how the geometrical coefficients evolve during this rotation for the first configuration in Fig.~\ref{fig1}(b).
Both $\gamma_0$ and $\gamma_2$ vary strongly and change signs, while $\gamma_1$, which vanishes at $\psi=0$ and $\pi/2$, becomes finite at intermediate angles.
This establishes that even at fixed microwave parameters, and therefore fixed $w_m$,
varying the orientation modifies the relative contributions of $m=0,1,2$,
which can change both the magnitude and sign of the three-body potential.
Microwave tuning and geometrical control thus provide complementary routes for engineering the nonadditive interaction.

By operating away from the compensation of first-order dipolar interactions,
the pairwise interaction can be suppressed so that the three-body interaction becomes dominant.
We tune the microwave parameters such that the full pairwise potential, containing first- and second-order contributions, vanishes at the lattice spacing, as illustrated in Fig.~\ref{fig4}(b). 
For the configuration in Fig.~\ref{fig4}(a), two pairwise interactions then vanish, leaving only the next-nearest-neighbor interaction at separation $\sqrt{2}a$, while a strong three-body contribution remains.

Figure~\ref{fig4}(c) shows that, for the configuration in Fig.~\ref{fig4}(a) for KAg, $a=532,\mathrm{nm}$, and $\Omega_{\rm tot}=1,\mathrm{MHz}$, the resulting three-body interaction $W$ exceeds the residual pairwise interaction $U=U_{ij}(\sqrt{2}a)$ in magnitude over a broad range of microwave parameters. The cancellation at $R=a$ needs to balance an attractive first-order $R^{-3}$ contribution against the repulsive second-order $R^{-6}$ contribution, and does not follow the same universal scaling in Eq.~\eqref{wi} because the two terms depend differently on $\Omega_{\rm tot}$. At $R=\sqrt{2}a$, the attractive term dominates due to its slower decay, making the residual pairwise interaction attractive and reversing the sign pattern of $W/U$ compared with Fig.~\ref{fig2}(d). 
These results demonstrate that double microwave dressing can suppress pairwise interactions while retaining a sizable three-body contribution, realizing a three-body-dominated regime compatible with collisional shielding.

In conclusion, we have shown that nonadditive three-body interactions are naturally induced by microwave shielding, required for collisional stabilization of ultracold molecules.
The microwave parameters and orientation flexibly control their magnitude, sign, and geometrical dependence.
For ultrapolar molecules in typical optical lattices, the three-body coefficients can reach the ten kHz scale, with interactions becoming comparable to pairwise contributions at dipolar compensation and even dominating when nearest-neighbor pairwise interactions are further suppressed.
These three-body interactions should be accounted for in quantum simulation with microwave shielded molecules,
even when targeting simulation of pairwise interactions.
At dipolar compensation, strong on-site repulsion can suppress double occupancy \cite{li:2026hubbard,perez:2026hubbard,stewartwiese:2026hubbard},
while the pairwise off-site interaction can be made isotropic in the plane of a two-dimensional lattice.
The three-body interaction, however, remains sensitive to the geometrical arrangement of occupied sites.
In a partially filled lattice, the three-body interaction can hence favor particular occupation patterns and stabilize unconventional ordered phases~\cite{bonnes:2010,bonnes:2011b}.

\textit{Acknowledgments---}
This work was supported by NWO VIDI (grant ID 10.61686/AKJWK33335).

\bibliography{ref}

\begin{thebibliography}{50}%
\makeatletter
\providecommand \@ifxundefined [1]{%
 \@ifx{#1\undefined}
}%
\providecommand \@ifnum [1]{%
 \ifnum #1\expandafter \@firstoftwo
 \else \expandafter \@secondoftwo
 \fi
}%
\providecommand \@ifx [1]{%
 \ifx #1\expandafter \@firstoftwo
 \else \expandafter \@secondoftwo
 \fi
}%
\providecommand \natexlab [1]{#1}%
\providecommand \enquote  [1]{``#1''}%
\providecommand \bibnamefont  [1]{#1}%
\providecommand \bibfnamefont [1]{#1}%
\providecommand \citenamefont [1]{#1}%
\providecommand \href@noop [0]{\@secondoftwo}%
\providecommand \href [0]{\begingroup \@sanitize@url \@href}%
\providecommand \@href[1]{\@@startlink{#1}\@@href}%
\providecommand \@@href[1]{\endgroup#1\@@endlink}%
\providecommand \@sanitize@url [0]{\catcode `\\12\catcode `\$12\catcode `\&12\catcode `\#12\catcode `\^12\catcode `\_12\catcode `\%12\relax}%
\providecommand \@@startlink[1]{}%
\providecommand \@@endlink[0]{}%
\providecommand \url  [0]{\begingroup\@sanitize@url \@url }%
\providecommand \@url [1]{\endgroup\@href {#1}{\urlprefix }}%
\providecommand \urlprefix  [0]{URL }%
\providecommand \Eprint [0]{\href }%
\providecommand \doibase [0]{http://dx.doi.org/}%
\providecommand \selectlanguage [0]{\@gobble}%
\providecommand \bibinfo  [0]{\@secondoftwo}%
\providecommand \bibfield  [0]{\@secondoftwo}%
\providecommand \translation [1]{[#1]}%
\providecommand \BibitemOpen [0]{}%
\providecommand \bibitemStop [0]{}%
\providecommand \bibitemNoStop [0]{.\EOS\space}%
\providecommand \EOS [0]{\spacefactor3000\relax}%
\providecommand \BibitemShut  [1]{\csname bibitem#1\endcsname}%
\let\auto@bib@innerbib\@empty
\bibitem [{\citenamefont {Gadway}\ and\ \citenamefont {Yan}(2016)}]{gadway:2016}%
  \BibitemOpen
  \bibfield  {author} {\bibinfo {author} {\bibfnamefont {B.}~\bibnamefont {Gadway}}\ and\ \bibinfo {author} {\bibfnamefont {B.}~\bibnamefont {Yan}},\ }\href {\doibase 10.1088/0953-4075/49/15/152002} {\bibfield  {journal} {\bibinfo  {journal} {J. Phys. B}\ }\textbf {\bibinfo {volume} {49}},\ \bibinfo {pages} {152002} (\bibinfo {year} {2016})}\BibitemShut {NoStop}%
\bibitem [{\citenamefont {Bohn}\ \emph {et~al.}(2017)\citenamefont {Bohn}, \citenamefont {Rey},\ and\ \citenamefont {Ye}}]{bohn:2017}%
  \BibitemOpen
  \bibfield  {author} {\bibinfo {author} {\bibfnamefont {J.~L.}\ \bibnamefont {Bohn}}, \bibinfo {author} {\bibfnamefont {A.~M.}\ \bibnamefont {Rey}}, \ and\ \bibinfo {author} {\bibfnamefont {J.}~\bibnamefont {Ye}},\ }\href {\doibase 10.1126/science.aam6299} {\bibfield  {journal} {\bibinfo  {journal} {Science}\ }\textbf {\bibinfo {volume} {357}},\ \bibinfo {pages} {1002} (\bibinfo {year} {2017})}\BibitemShut {NoStop}%
\bibitem [{\citenamefont {Cornish}\ \emph {et~al.}(2024)\citenamefont {Cornish}, \citenamefont {Tarbutt},\ and\ \citenamefont {Hazzard}}]{cornish:2024}%
  \BibitemOpen
  \bibfield  {author} {\bibinfo {author} {\bibfnamefont {S.~L.}\ \bibnamefont {Cornish}}, \bibinfo {author} {\bibfnamefont {M.~R.}\ \bibnamefont {Tarbutt}}, \ and\ \bibinfo {author} {\bibfnamefont {K.~R.~A.}\ \bibnamefont {Hazzard}},\ }\href {\doibase 10.1038/s41567-024-02453-9} {\bibfield  {journal} {\bibinfo  {journal} {Nat. Phys.}\ }\textbf {\bibinfo {volume} {20}},\ \bibinfo {pages} {730} (\bibinfo {year} {2024})}\BibitemShut {NoStop}%
\bibitem [{\citenamefont {Langen}\ \emph {et~al.}(2024)\citenamefont {Langen}, \citenamefont {Valtolina}, \citenamefont {Wang},\ and\ \citenamefont {Ye}}]{langen:2024}%
  \BibitemOpen
  \bibfield  {author} {\bibinfo {author} {\bibfnamefont {T.}~\bibnamefont {Langen}}, \bibinfo {author} {\bibfnamefont {G.}~\bibnamefont {Valtolina}}, \bibinfo {author} {\bibfnamefont {D.}~\bibnamefont {Wang}}, \ and\ \bibinfo {author} {\bibfnamefont {J.}~\bibnamefont {Ye}},\ }\href {\doibase 10.1038/s41567-024-02423-1} {\bibfield  {journal} {\bibinfo  {journal} {Nat. Phys.}\ }\textbf {\bibinfo {volume} {20}},\ \bibinfo {pages} {702} (\bibinfo {year} {2024})}\BibitemShut {NoStop}%
\bibitem [{\citenamefont {Ni}\ \emph {et~al.}(2008)\citenamefont {Ni}, \citenamefont {Ospelkaus}, \citenamefont {de~Miranda}, \citenamefont {Pe'er}, \citenamefont {Neyenhuis}, \citenamefont {Zirbel}, \citenamefont {Kotochigova}, \citenamefont {Julienne}, \citenamefont {Jin},\ and\ \citenamefont {Ye}}]{ni:2008}%
  \BibitemOpen
  \bibfield  {author} {\bibinfo {author} {\bibfnamefont {K.-K.}\ \bibnamefont {Ni}}, \bibinfo {author} {\bibfnamefont {S.}~\bibnamefont {Ospelkaus}}, \bibinfo {author} {\bibfnamefont {M.~H.~G.}\ \bibnamefont {de~Miranda}}, \bibinfo {author} {\bibfnamefont {A.}~\bibnamefont {Pe'er}}, \bibinfo {author} {\bibfnamefont {B.}~\bibnamefont {Neyenhuis}}, \bibinfo {author} {\bibfnamefont {J.~J.}\ \bibnamefont {Zirbel}}, \bibinfo {author} {\bibfnamefont {S.}~\bibnamefont {Kotochigova}}, \bibinfo {author} {\bibfnamefont {P.~S.}\ \bibnamefont {Julienne}}, \bibinfo {author} {\bibfnamefont {D.~S.}\ \bibnamefont {Jin}}, \ and\ \bibinfo {author} {\bibfnamefont {J.}~\bibnamefont {Ye}},\ }\href {\doibase 10.1126/science.1163861} {\bibfield  {journal} {\bibinfo  {journal} {Science}\ }\textbf {\bibinfo {volume} {322}},\ \bibinfo {pages} {231} (\bibinfo {year} {2008})}\BibitemShut {NoStop}%
\bibitem [{\citenamefont {Park}\ \emph {et~al.}(2015)\citenamefont {Park}, \citenamefont {Will},\ and\ \citenamefont {Zwierlein}}]{park:2015}%
  \BibitemOpen
  \bibfield  {author} {\bibinfo {author} {\bibfnamefont {J.~W.}\ \bibnamefont {Park}}, \bibinfo {author} {\bibfnamefont {S.~A.}\ \bibnamefont {Will}}, \ and\ \bibinfo {author} {\bibfnamefont {M.~W.}\ \bibnamefont {Zwierlein}},\ }\href {\doibase 10.1103/PhysRevLett.114.205302} {\bibfield  {journal} {\bibinfo  {journal} {Phys. Rev. Lett.}\ }\textbf {\bibinfo {volume} {114}},\ \bibinfo {pages} {205302} (\bibinfo {year} {2015})}\BibitemShut {NoStop}%
\bibitem [{\citenamefont {Zhang}\ \emph {et~al.}(2021)\citenamefont {Zhang}, \citenamefont {Chen}, \citenamefont {Yao},\ and\ \citenamefont {Chin}}]{zhang:2021}%
  \BibitemOpen
  \bibfield  {author} {\bibinfo {author} {\bibfnamefont {Z.}~\bibnamefont {Zhang}}, \bibinfo {author} {\bibfnamefont {L.}~\bibnamefont {Chen}}, \bibinfo {author} {\bibfnamefont {K.-X.}\ \bibnamefont {Yao}}, \ and\ \bibinfo {author} {\bibfnamefont {C.}~\bibnamefont {Chin}},\ }\href {\doibase 10.1038/s41586-021-03443-0} {\bibfield  {journal} {\bibinfo  {journal} {Nature}\ }\textbf {\bibinfo {volume} {592}},\ \bibinfo {pages} {708} (\bibinfo {year} {2021})}\BibitemShut {NoStop}%
\bibitem [{\citenamefont {De~Marco}\ \emph {et~al.}(2019)\citenamefont {De~Marco}, \citenamefont {Valtolina}, \citenamefont {Matsuda}, \citenamefont {Tobias}, \citenamefont {Covey},\ and\ \citenamefont {Ye}}]{demarco:2019}%
  \BibitemOpen
  \bibfield  {author} {\bibinfo {author} {\bibfnamefont {L.}~\bibnamefont {De~Marco}}, \bibinfo {author} {\bibfnamefont {G.}~\bibnamefont {Valtolina}}, \bibinfo {author} {\bibfnamefont {K.}~\bibnamefont {Matsuda}}, \bibinfo {author} {\bibfnamefont {W.~G.}\ \bibnamefont {Tobias}}, \bibinfo {author} {\bibfnamefont {J.~P.}\ \bibnamefont {Covey}}, \ and\ \bibinfo {author} {\bibfnamefont {J.}~\bibnamefont {Ye}},\ }\href {\doibase 10.1126/science.aau7230} {\bibfield  {journal} {\bibinfo  {journal} {Science}\ }\textbf {\bibinfo {volume} {363}},\ \bibinfo {pages} {853} (\bibinfo {year} {2019})}\BibitemShut {NoStop}%
\bibitem [{\citenamefont {Karman}\ and\ \citenamefont {Hutson}(2018)}]{karman:2018}%
  \BibitemOpen
  \bibfield  {author} {\bibinfo {author} {\bibfnamefont {T.}~\bibnamefont {Karman}}\ and\ \bibinfo {author} {\bibfnamefont {J.~M.}\ \bibnamefont {Hutson}},\ }\href {\doibase 10.1103/PhysRevLett.121.163401} {\bibfield  {journal} {\bibinfo  {journal} {Phys. Rev. Lett.}\ }\textbf {\bibinfo {volume} {121}},\ \bibinfo {pages} {163401} (\bibinfo {year} {2018})}\BibitemShut {NoStop}%
\bibitem [{\citenamefont {Bigagli}\ \emph {et~al.}(2023)\citenamefont {Bigagli}, \citenamefont {Warner}, \citenamefont {Yuan}, \citenamefont {Zhang}, \citenamefont {Stevenson}, \citenamefont {Karman},\ and\ \citenamefont {Will}}]{bigagli:2023}%
  \BibitemOpen
  \bibfield  {author} {\bibinfo {author} {\bibfnamefont {N.}~\bibnamefont {Bigagli}}, \bibinfo {author} {\bibfnamefont {C.}~\bibnamefont {Warner}}, \bibinfo {author} {\bibfnamefont {W.}~\bibnamefont {Yuan}}, \bibinfo {author} {\bibfnamefont {S.}~\bibnamefont {Zhang}}, \bibinfo {author} {\bibfnamefont {I.}~\bibnamefont {Stevenson}}, \bibinfo {author} {\bibfnamefont {T.}~\bibnamefont {Karman}}, \ and\ \bibinfo {author} {\bibfnamefont {S.}~\bibnamefont {Will}},\ }\href {\doibase 10.1038/s41567-023-02200-6} {\bibfield  {journal} {\bibinfo  {journal} {Nat. Phys.}\ }\textbf {\bibinfo {volume} {19}},\ \bibinfo {pages} {1579} (\bibinfo {year} {2023})}\BibitemShut {NoStop}%
\bibitem [{\citenamefont {Lin}\ \emph {et~al.}(2023)\citenamefont {Lin}, \citenamefont {Chen}, \citenamefont {Jin}, \citenamefont {Shi}, \citenamefont {Deng}, \citenamefont {Zhang}, \citenamefont {Qu{\'e}m{\'e}ner}, \citenamefont {Shi}, \citenamefont {Yi},\ and\ \citenamefont {Wang}}]{lin:2023}%
  \BibitemOpen
  \bibfield  {author} {\bibinfo {author} {\bibfnamefont {J.}~\bibnamefont {Lin}}, \bibinfo {author} {\bibfnamefont {G.}~\bibnamefont {Chen}}, \bibinfo {author} {\bibfnamefont {M.}~\bibnamefont {Jin}}, \bibinfo {author} {\bibfnamefont {Z.}~\bibnamefont {Shi}}, \bibinfo {author} {\bibfnamefont {F.}~\bibnamefont {Deng}}, \bibinfo {author} {\bibfnamefont {W.}~\bibnamefont {Zhang}}, \bibinfo {author} {\bibfnamefont {G.}~\bibnamefont {Qu{\'e}m{\'e}ner}}, \bibinfo {author} {\bibfnamefont {T.}~\bibnamefont {Shi}}, \bibinfo {author} {\bibfnamefont {S.}~\bibnamefont {Yi}}, \ and\ \bibinfo {author} {\bibfnamefont {D.}~\bibnamefont {Wang}},\ }\href {\doibase 10.1103/PhysRevX.13.031032} {\bibfield  {journal} {\bibinfo  {journal} {Phys. Rev. X}\ }\textbf {\bibinfo {volume} {13}},\ \bibinfo {pages} {031032} (\bibinfo {year} {2023})}\BibitemShut {NoStop}%
\bibitem [{\citenamefont {Yuan}\ \emph {et~al.}(2026)\citenamefont {Yuan}, \citenamefont {Zhang}, \citenamefont {Bigagli}, \citenamefont {Kwak}, \citenamefont {Warner}, \citenamefont {Karman}, \citenamefont {Stevenson},\ and\ \citenamefont {Will}}]{yuan2025extreme}%
  \BibitemOpen
  \bibfield  {author} {\bibinfo {author} {\bibfnamefont {W.}~\bibnamefont {Yuan}}, \bibinfo {author} {\bibfnamefont {S.}~\bibnamefont {Zhang}}, \bibinfo {author} {\bibfnamefont {N.}~\bibnamefont {Bigagli}}, \bibinfo {author} {\bibfnamefont {H.}~\bibnamefont {Kwak}}, \bibinfo {author} {\bibfnamefont {C.}~\bibnamefont {Warner}}, \bibinfo {author} {\bibfnamefont {T.}~\bibnamefont {Karman}}, \bibinfo {author} {\bibfnamefont {I.}~\bibnamefont {Stevenson}}, \ and\ \bibinfo {author} {\bibfnamefont {S.}~\bibnamefont {Will}},\ }\href {\doibase 10.1126/science.adz0521} {\bibfield  {journal} {\bibinfo  {journal} {Science}\ }\textbf {\bibinfo {volume} {393}},\ \bibinfo {pages} {1226} (\bibinfo {year} {2026})}\BibitemShut {NoStop}%
\bibitem [{\citenamefont {Schindewolf}\ \emph {et~al.}(2022)\citenamefont {Schindewolf}, \citenamefont {Bause}, \citenamefont {Chen}, \citenamefont {Duda}, \citenamefont {Karman}, \citenamefont {Bloch},\ and\ \citenamefont {Luo}}]{schindewolf:2022}%
  \BibitemOpen
  \bibfield  {author} {\bibinfo {author} {\bibfnamefont {A.}~\bibnamefont {Schindewolf}}, \bibinfo {author} {\bibfnamefont {R.}~\bibnamefont {Bause}}, \bibinfo {author} {\bibfnamefont {X.-Y.}\ \bibnamefont {Chen}}, \bibinfo {author} {\bibfnamefont {M.}~\bibnamefont {Duda}}, \bibinfo {author} {\bibfnamefont {T.}~\bibnamefont {Karman}}, \bibinfo {author} {\bibfnamefont {I.}~\bibnamefont {Bloch}}, \ and\ \bibinfo {author} {\bibfnamefont {X.-Y.}\ \bibnamefont {Luo}},\ }\href {\doibase 10.1038/s41586-022-04900-0} {\bibfield  {journal} {\bibinfo  {journal} {Nature}\ }\textbf {\bibinfo {volume} {607}},\ \bibinfo {pages} {677} (\bibinfo {year} {2022})}\BibitemShut {NoStop}%
\bibitem [{\citenamefont {Shi}\ \emph {et~al.}(2026)\citenamefont {Shi}, \citenamefont {Huang}, \citenamefont {Deng}, \citenamefont {Jin}, \citenamefont {Yi}, \citenamefont {Shi},\ and\ \citenamefont {Wang}}]{shi:2025}%
  \BibitemOpen
  \bibfield  {author} {\bibinfo {author} {\bibfnamefont {Z.}~\bibnamefont {Shi}}, \bibinfo {author} {\bibfnamefont {Z.}~\bibnamefont {Huang}}, \bibinfo {author} {\bibfnamefont {F.}~\bibnamefont {Deng}}, \bibinfo {author} {\bibfnamefont {W.-J.}\ \bibnamefont {Jin}}, \bibinfo {author} {\bibfnamefont {S.}~\bibnamefont {Yi}}, \bibinfo {author} {\bibfnamefont {T.}~\bibnamefont {Shi}}, \ and\ \bibinfo {author} {\bibfnamefont {D.}~\bibnamefont {Wang}},\ }\href {\doibase 10.1038/s41567-026-03362-9} {\bibfield  {journal} {\bibinfo  {journal} {Nat. Phys.}\ }\textbf {\bibinfo {volume} {22}},\ \bibinfo {pages} {1467} (\bibinfo {year} {2026})}\BibitemShut {NoStop}%
\bibitem [{\citenamefont {Deng}\ \emph {et~al.}(2023)\citenamefont {Deng}, \citenamefont {Chen}, \citenamefont {Luo}, \citenamefont {Zhang}, \citenamefont {Yi},\ and\ \citenamefont {Shi}}]{deng:2023}%
  \BibitemOpen
  \bibfield  {author} {\bibinfo {author} {\bibfnamefont {F.}~\bibnamefont {Deng}}, \bibinfo {author} {\bibfnamefont {X.-Y.}\ \bibnamefont {Chen}}, \bibinfo {author} {\bibfnamefont {X.-Y.}\ \bibnamefont {Luo}}, \bibinfo {author} {\bibfnamefont {W.}~\bibnamefont {Zhang}}, \bibinfo {author} {\bibfnamefont {S.}~\bibnamefont {Yi}}, \ and\ \bibinfo {author} {\bibfnamefont {T.}~\bibnamefont {Shi}},\ }\href {\doibase 10.1103/PhysRevLett.130.183001} {\bibfield  {journal} {\bibinfo  {journal} {Phys. Rev. Lett.}\ }\textbf {\bibinfo {volume} {130}},\ \bibinfo {pages} {183001} (\bibinfo {year} {2023})}\BibitemShut {NoStop}%
\bibitem [{\citenamefont {Karman}\ \emph {et~al.}(2025)\citenamefont {Karman}, \citenamefont {Bigagli}, \citenamefont {Yuan}, \citenamefont {Zhang}, \citenamefont {Stevenson},\ and\ \citenamefont {Will}}]{karman:2025}%
  \BibitemOpen
  \bibfield  {author} {\bibinfo {author} {\bibfnamefont {T.}~\bibnamefont {Karman}}, \bibinfo {author} {\bibfnamefont {N.}~\bibnamefont {Bigagli}}, \bibinfo {author} {\bibfnamefont {W.}~\bibnamefont {Yuan}}, \bibinfo {author} {\bibfnamefont {S.}~\bibnamefont {Zhang}}, \bibinfo {author} {\bibfnamefont {I.}~\bibnamefont {Stevenson}}, \ and\ \bibinfo {author} {\bibfnamefont {S.}~\bibnamefont {Will}},\ }\href {\doibase 10.1103/b8pm-3prn} {\bibfield  {journal} {\bibinfo  {journal} {PRX Quantum}\ }\textbf {\bibinfo {volume} {6}},\ \bibinfo {pages} {020358} (\bibinfo {year} {2025})}\BibitemShut {NoStop}%
\bibitem [{\citenamefont {Bigagli}\ \emph {et~al.}(2024)\citenamefont {Bigagli}, \citenamefont {Yuan}, \citenamefont {Zhang}, \citenamefont {Bulatovic}, \citenamefont {Karman}, \citenamefont {Stevenson},\ and\ \citenamefont {Will}}]{bigagli:2024}%
  \BibitemOpen
  \bibfield  {author} {\bibinfo {author} {\bibfnamefont {N.}~\bibnamefont {Bigagli}}, \bibinfo {author} {\bibfnamefont {W.}~\bibnamefont {Yuan}}, \bibinfo {author} {\bibfnamefont {S.}~\bibnamefont {Zhang}}, \bibinfo {author} {\bibfnamefont {B.}~\bibnamefont {Bulatovic}}, \bibinfo {author} {\bibfnamefont {T.}~\bibnamefont {Karman}}, \bibinfo {author} {\bibfnamefont {I.}~\bibnamefont {Stevenson}}, \ and\ \bibinfo {author} {\bibfnamefont {S.}~\bibnamefont {Will}},\ }\href {\doibase 10.1038/s41586-024-07492-z} {\bibfield  {journal} {\bibinfo  {journal} {Nature}\ }\textbf {\bibinfo {volume} {631}},\ \bibinfo {pages} {289} (\bibinfo {year} {2024})}\BibitemShut {NoStop}%
\bibitem [{\citenamefont {Zhang}\ \emph {et~al.}(2026)\citenamefont {Zhang}, \citenamefont {Yuan}, \citenamefont {Bigagli}, \citenamefont {Kwak}, \citenamefont {Karman}, \citenamefont {Stevenson},\ and\ \citenamefont {Will}}]{zhang:2026}%
  \BibitemOpen
  \bibfield  {author} {\bibinfo {author} {\bibfnamefont {S.}~\bibnamefont {Zhang}}, \bibinfo {author} {\bibfnamefont {W.}~\bibnamefont {Yuan}}, \bibinfo {author} {\bibfnamefont {N.}~\bibnamefont {Bigagli}}, \bibinfo {author} {\bibfnamefont {H.}~\bibnamefont {Kwak}}, \bibinfo {author} {\bibfnamefont {T.}~\bibnamefont {Karman}}, \bibinfo {author} {\bibfnamefont {I.}~\bibnamefont {Stevenson}}, \ and\ \bibinfo {author} {\bibfnamefont {S.}~\bibnamefont {Will}},\ }\href {\doibase 10.1038/s41586-026-10245-9} {\bibfield  {journal} {\bibinfo  {journal} {Nature}\ }\textbf {\bibinfo {volume} {651}},\ \bibinfo {pages} {601} (\bibinfo {year} {2026})}\BibitemShut {NoStop}%
\bibitem [{\citenamefont {Xie}\ \emph {et~al.}(2020)\citenamefont {Xie}, \citenamefont {Lepers}, \citenamefont {Vexiau}, \citenamefont {Orb{\'a}n}, \citenamefont {Dulieu},\ and\ \citenamefont {Bouloufa-Maafa}}]{xie.T}%
  \BibitemOpen
  \bibfield  {author} {\bibinfo {author} {\bibfnamefont {T.}~\bibnamefont {Xie}}, \bibinfo {author} {\bibfnamefont {M.}~\bibnamefont {Lepers}}, \bibinfo {author} {\bibfnamefont {R.}~\bibnamefont {Vexiau}}, \bibinfo {author} {\bibfnamefont {A.}~\bibnamefont {Orb{\'a}n}}, \bibinfo {author} {\bibfnamefont {O.}~\bibnamefont {Dulieu}}, \ and\ \bibinfo {author} {\bibfnamefont {N.}~\bibnamefont {Bouloufa-Maafa}},\ }\href {\doibase 10.1103/PhysRevLett.125.153202} {\bibfield  {journal} {\bibinfo  {journal} {Phys. Rev. Lett.}\ }\textbf {\bibinfo {volume} {125}},\ \bibinfo {pages} {153202} (\bibinfo {year} {2020})}\BibitemShut {NoStop}%
\bibitem [{\citenamefont {Karam}\ \emph {et~al.}(2023)\citenamefont {Karam}, \citenamefont {Vexiau}, \citenamefont {Bouloufa-Maafa}, \citenamefont {Dulieu}, \citenamefont {Lepers}, \citenamefont {Borgloh}, \citenamefont {Ospelkaus},\ and\ \citenamefont {Karpa}}]{karam}%
  \BibitemOpen
  \bibfield  {author} {\bibinfo {author} {\bibfnamefont {C.}~\bibnamefont {Karam}}, \bibinfo {author} {\bibfnamefont {R.}~\bibnamefont {Vexiau}}, \bibinfo {author} {\bibfnamefont {N.}~\bibnamefont {Bouloufa-Maafa}}, \bibinfo {author} {\bibfnamefont {O.}~\bibnamefont {Dulieu}}, \bibinfo {author} {\bibfnamefont {M.}~\bibnamefont {Lepers}}, \bibinfo {author} {\bibfnamefont {M.~M. z.~A.}\ \bibnamefont {Borgloh}}, \bibinfo {author} {\bibfnamefont {S.}~\bibnamefont {Ospelkaus}}, \ and\ \bibinfo {author} {\bibfnamefont {L.}~\bibnamefont {Karpa}},\ }\href {\doibase 10.1103/PhysRevResearch.5.033074} {\bibfield  {journal} {\bibinfo  {journal} {Phys. Rev. Res.}\ }\textbf {\bibinfo {volume} {5}},\ \bibinfo {pages} {033074} (\bibinfo {year} {2023})}\BibitemShut {NoStop}%
\bibitem [{\citenamefont {Augustovičová}\ and\ \citenamefont {Bohn}(2019)}]{augustovicova}%
  \BibitemOpen
  \bibfield  {author} {\bibinfo {author} {\bibfnamefont {L.~D.}\ \bibnamefont {Augustovičová}}\ and\ \bibinfo {author} {\bibfnamefont {J.~L.}\ \bibnamefont {Bohn}},\ }\href {\doibase 10.1088/1367-2630/ab4720} {\bibfield  {journal} {\bibinfo  {journal} {New Journal of Physics}\ }\textbf {\bibinfo {volume} {21}},\ \bibinfo {pages} {103022} (\bibinfo {year} {2019})}\BibitemShut {NoStop}%
\bibitem [{\citenamefont {Gonz{\'a}lez-Mart{\'{\i}}nez}\ \emph {et~al.}(2017)\citenamefont {Gonz{\'a}lez-Mart{\'{\i}}nez}, \citenamefont {Bohn},\ and\ \citenamefont {Qu{\'e}m{\'e}ner}}]{gonzalez}%
  \BibitemOpen
  \bibfield  {author} {\bibinfo {author} {\bibfnamefont {M.~L.}\ \bibnamefont {Gonz{\'a}lez-Mart{\'{\i}}nez}}, \bibinfo {author} {\bibfnamefont {J.~L.}\ \bibnamefont {Bohn}}, \ and\ \bibinfo {author} {\bibfnamefont {G.}~\bibnamefont {Qu{\'e}m{\'e}ner}},\ }\href {\doibase 10.1103/PhysRevA.96.032718} {\bibfield  {journal} {\bibinfo  {journal} {Phys. Rev. A}\ }\textbf {\bibinfo {volume} {96}},\ \bibinfo {pages} {032718} (\bibinfo {year} {2017})}\BibitemShut {NoStop}%
\bibitem [{\citenamefont {Matsuda}\ \emph {et~al.}(2020)\citenamefont {Matsuda}, \citenamefont {De~Marco}, \citenamefont {Li}, \citenamefont {Tobias}, \citenamefont {Valtolina}, \citenamefont {Qu{\'e}m{\'e}ner},\ and\ \citenamefont {Ye}}]{matsuda2020resonant}%
  \BibitemOpen
  \bibfield  {author} {\bibinfo {author} {\bibfnamefont {K.}~\bibnamefont {Matsuda}}, \bibinfo {author} {\bibfnamefont {L.}~\bibnamefont {De~Marco}}, \bibinfo {author} {\bibfnamefont {J.-R.}\ \bibnamefont {Li}}, \bibinfo {author} {\bibfnamefont {W.~G.}\ \bibnamefont {Tobias}}, \bibinfo {author} {\bibfnamefont {G.}~\bibnamefont {Valtolina}}, \bibinfo {author} {\bibfnamefont {G.}~\bibnamefont {Qu{\'e}m{\'e}ner}}, \ and\ \bibinfo {author} {\bibfnamefont {J.}~\bibnamefont {Ye}},\ }\href@noop {} {\bibfield  {journal} {\bibinfo  {journal} {Science}\ }\textbf {\bibinfo {volume} {370}},\ \bibinfo {pages} {1324} (\bibinfo {year} {2020})}\BibitemShut {NoStop}%
\bibitem [{\citenamefont {Lin}\ \emph {et~al.}(2026)\citenamefont {Lin}, \citenamefont {Carroll}, \citenamefont {Martin}, \citenamefont {Miller}, \citenamefont {Wang}, \citenamefont {Xu}, \citenamefont {Bohn}, \citenamefont {de~Jongh},\ and\ \citenamefont {Ye}}]{ye:2026}%
  \BibitemOpen
  \bibfield  {author} {\bibinfo {author} {\bibfnamefont {J.}~\bibnamefont {Lin}}, \bibinfo {author} {\bibfnamefont {A.~N.}\ \bibnamefont {Carroll}}, \bibinfo {author} {\bibfnamefont {P.}~\bibnamefont {Martin}}, \bibinfo {author} {\bibfnamefont {C.}~\bibnamefont {Miller}}, \bibinfo {author} {\bibfnamefont {R.~R.~W.}\ \bibnamefont {Wang}}, \bibinfo {author} {\bibfnamefont {K.}~\bibnamefont {Xu}}, \bibinfo {author} {\bibfnamefont {J.~L.}\ \bibnamefont {Bohn}}, \bibinfo {author} {\bibfnamefont {T.}~\bibnamefont {de~Jongh}}, \ and\ \bibinfo {author} {\bibfnamefont {J.}~\bibnamefont {Ye}},\ }\href {https://arxiv.org/abs/2606.14135} {\enquote {\bibinfo {title} {Fermi gas of polar molecules in the pauli-blocked regime},}\ } (\bibinfo {year} {2026}),\ \Eprint {http://arxiv.org/abs/2606.14135} {arXiv:2606.14135 [cond-mat.quant-gas]} \BibitemShut {NoStop}%
\bibitem [{\citenamefont {Jung}\ \emph {et~al.}(2026)\citenamefont {Jung}, \citenamefont {Futamura}, \citenamefont {Ciardi}, \citenamefont {Rosenberg}, \citenamefont {Alaoui}, \citenamefont {Lu}, \citenamefont {Pohl},\ and\ \citenamefont {Bakr}}]{bakr:2026}%
  \BibitemOpen
  \bibfield  {author} {\bibinfo {author} {\bibfnamefont {J.}~\bibnamefont {Jung}}, \bibinfo {author} {\bibfnamefont {L.}~\bibnamefont {Futamura}}, \bibinfo {author} {\bibfnamefont {M.}~\bibnamefont {Ciardi}}, \bibinfo {author} {\bibfnamefont {J.}~\bibnamefont {Rosenberg}}, \bibinfo {author} {\bibfnamefont {Y.~A.}\ \bibnamefont {Alaoui}}, \bibinfo {author} {\bibfnamefont {Y.}~\bibnamefont {Lu}}, \bibinfo {author} {\bibfnamefont {T.}~\bibnamefont {Pohl}}, \ and\ \bibinfo {author} {\bibfnamefont {W.~S.}\ \bibnamefont {Bakr}},\ }\href {https://arxiv.org/abs/2608.31116} {\enquote {\bibinfo {title} {Static-field shielding of bosonic molecules: Evaporation to degeneracy and self-bound droplets},}\ } (\bibinfo {year} {2026}),\ \Eprint {http://arxiv.org/abs/2608.31116} {arXiv:2608.31116 [cond-mat.quant-gas]} \BibitemShut {NoStop}%
\bibitem [{\citenamefont {Karman}\ \emph {et~al.}(2026)\citenamefont {Karman}, \citenamefont {Will},\ and\ \citenamefont {Yan}}]{karman:2026blockade}%
  \BibitemOpen
  \bibfield  {author} {\bibinfo {author} {\bibfnamefont {T.}~\bibnamefont {Karman}}, \bibinfo {author} {\bibfnamefont {S.}~\bibnamefont {Will}}, \ and\ \bibinfo {author} {\bibfnamefont {Z.~Z.}\ \bibnamefont {Yan}},\ }\href {\doibase 10.1103/vprz-jqz5} {\bibfield  {journal} {\bibinfo  {journal} {Phys. Rev. Lett.}\ }\textbf {\bibinfo {volume} {137}},\ \bibinfo {pages} {113402} (\bibinfo {year} {2026})}\BibitemShut {NoStop}%
\bibitem [{\citenamefont {Dutta}\ \emph {et~al.}(2015)\citenamefont {Dutta}, \citenamefont {Gajda}, \citenamefont {Hauke}, \citenamefont {Lewenstein}, \citenamefont {Lühmann}, \citenamefont {Malomed}, \citenamefont {Sowiński},\ and\ \citenamefont {Zakrzewski}}]{Dutta_2015}%
  \BibitemOpen
  \bibfield  {author} {\bibinfo {author} {\bibfnamefont {O.}~\bibnamefont {Dutta}}, \bibinfo {author} {\bibfnamefont {M.}~\bibnamefont {Gajda}}, \bibinfo {author} {\bibfnamefont {P.}~\bibnamefont {Hauke}}, \bibinfo {author} {\bibfnamefont {M.}~\bibnamefont {Lewenstein}}, \bibinfo {author} {\bibfnamefont {D.-S.}\ \bibnamefont {Lühmann}}, \bibinfo {author} {\bibfnamefont {B.~A.}\ \bibnamefont {Malomed}}, \bibinfo {author} {\bibfnamefont {T.}~\bibnamefont {Sowiński}}, \ and\ \bibinfo {author} {\bibfnamefont {J.}~\bibnamefont {Zakrzewski}},\ }\href {\doibase 10.1088/0034-4885/78/6/066001} {\bibfield  {journal} {\bibinfo  {journal} {Reports on Progress in Physics}\ }\textbf {\bibinfo {volume} {78}},\ \bibinfo {pages} {066001} (\bibinfo {year} {2015})}\BibitemShut {NoStop}%
\bibitem [{\citenamefont {Chanda}\ \emph {et~al.}(2025)\citenamefont {Chanda}, \citenamefont {Barbiero}, \citenamefont {Lewenstein}, \citenamefont {Mark},\ and\ \citenamefont {Zakrzewski}}]{Chanda_2025}%
  \BibitemOpen
  \bibfield  {author} {\bibinfo {author} {\bibfnamefont {T.}~\bibnamefont {Chanda}}, \bibinfo {author} {\bibfnamefont {L.}~\bibnamefont {Barbiero}}, \bibinfo {author} {\bibfnamefont {M.}~\bibnamefont {Lewenstein}}, \bibinfo {author} {\bibfnamefont {M.~J.}\ \bibnamefont {Mark}}, \ and\ \bibinfo {author} {\bibfnamefont {J.}~\bibnamefont {Zakrzewski}},\ }\href {\doibase 10.1088/1361-6633/adc3a7} {\bibfield  {journal} {\bibinfo  {journal} {Reports on Progress in Physics}\ }\textbf {\bibinfo {volume} {88}},\ \bibinfo {pages} {044501} (\bibinfo {year} {2025})}\BibitemShut {NoStop}%
\bibitem [{\citenamefont {Christakis}\ \emph {et~al.}(2023)\citenamefont {Christakis}, \citenamefont {Rosenberg}, \citenamefont {Raj}, \citenamefont {Chi}, \citenamefont {Morningstar}, \citenamefont {Huse}, \citenamefont {Yan},\ and\ \citenamefont {Bakr}}]{christakis:2023}%
  \BibitemOpen
  \bibfield  {author} {\bibinfo {author} {\bibfnamefont {L.}~\bibnamefont {Christakis}}, \bibinfo {author} {\bibfnamefont {J.~S.}\ \bibnamefont {Rosenberg}}, \bibinfo {author} {\bibfnamefont {R.}~\bibnamefont {Raj}}, \bibinfo {author} {\bibfnamefont {S.}~\bibnamefont {Chi}}, \bibinfo {author} {\bibfnamefont {A.}~\bibnamefont {Morningstar}}, \bibinfo {author} {\bibfnamefont {D.~A.}\ \bibnamefont {Huse}}, \bibinfo {author} {\bibfnamefont {Z.~Z.}\ \bibnamefont {Yan}}, \ and\ \bibinfo {author} {\bibfnamefont {W.~S.}\ \bibnamefont {Bakr}},\ }\href {\doibase 10.1038/s41586-022-05558-4} {\bibfield  {journal} {\bibinfo  {journal} {Nature}\ }\textbf {\bibinfo {volume} {614}},\ \bibinfo {pages} {64} (\bibinfo {year} {2023})}\BibitemShut {NoStop}%
\bibitem [{\citenamefont {Mortlock}\ \emph {et~al.}(2026)\citenamefont {Mortlock}, \citenamefont {Raghuram}, \citenamefont {Maddox}, \citenamefont {Gregory},\ and\ \citenamefont {Cornish}}]{mortlock:2026}%
  \BibitemOpen
  \bibfield  {author} {\bibinfo {author} {\bibfnamefont {J.~M.}\ \bibnamefont {Mortlock}}, \bibinfo {author} {\bibfnamefont {A.~P.}\ \bibnamefont {Raghuram}}, \bibinfo {author} {\bibfnamefont {B.~P.}\ \bibnamefont {Maddox}}, \bibinfo {author} {\bibfnamefont {P.~D.}\ \bibnamefont {Gregory}}, \ and\ \bibinfo {author} {\bibfnamefont {S.~L.}\ \bibnamefont {Cornish}},\ }\href {\doibase 10.1038/s41467-025-67212-7} {\bibfield  {journal} {\bibinfo  {journal} {Nat. Commun.}\ }\textbf {\bibinfo {volume} {17}},\ \bibinfo {pages} {518} (\bibinfo {year} {2026})}\BibitemShut {NoStop}%
\bibitem [{\citenamefont {Li}\ \emph {et~al.}(2026)\citenamefont {Li}, \citenamefont {Alhyder}, \citenamefont {Schindewolf}, \citenamefont {Hazzard}, \citenamefont {Lemeshko},\ and\ \citenamefont {Koutentakis}}]{li:2026hubbard}%
  \BibitemOpen
  \bibfield  {author} {\bibinfo {author} {\bibfnamefont {J.-L.}\ \bibnamefont {Li}}, \bibinfo {author} {\bibfnamefont {R.}~\bibnamefont {Alhyder}}, \bibinfo {author} {\bibfnamefont {A.}~\bibnamefont {Schindewolf}}, \bibinfo {author} {\bibfnamefont {K.~R.~A.}\ \bibnamefont {Hazzard}}, \bibinfo {author} {\bibfnamefont {M.}~\bibnamefont {Lemeshko}}, \ and\ \bibinfo {author} {\bibfnamefont {G.~M.}\ \bibnamefont {Koutentakis}},\ }\href {\doibase 10.48550/arXiv.2607.27107} {\enquote {\bibinfo {title} {{Engineering SU($N$)-Symmetric Hubbard Models with Microwave-Shielded Dipolar Molecules}},}\ } (\bibinfo {year} {2026}),\ \Eprint {http://arxiv.org/abs/2607.27107} {arXiv:2607.27107 [cond-mat.quant-gas]} \BibitemShut {NoStop}%
\bibitem [{\citenamefont {P{\'e}rez}\ \emph {et~al.}(2026)\citenamefont {P{\'e}rez}, \citenamefont {Desroches}, \citenamefont {Hazzard},\ and\ \citenamefont {Karman}}]{perez:2026hubbard}%
  \BibitemOpen
  \bibfield  {author} {\bibinfo {author} {\bibfnamefont {K.}~\bibnamefont {P{\'e}rez}}, \bibinfo {author} {\bibfnamefont {J.~W.}\ \bibnamefont {Desroches}}, \bibinfo {author} {\bibfnamefont {K.~R.~A.}\ \bibnamefont {Hazzard}}, \ and\ \bibinfo {author} {\bibfnamefont {T.}~\bibnamefont {Karman}},\ }\href {\doibase 10.48550/arXiv.2609.20286} {\enquote {\bibinfo {title} {{Hubbard models with ultracold shielded molecules in optical lattices}},}\ } (\bibinfo {year} {2026}),\ \Eprint {http://arxiv.org/abs/2609.20286} {arXiv:2609.20286 [cond-mat.quant-gas]} \BibitemShut {NoStop}%
\bibitem [{\citenamefont {Stewart-Wiese}\ \emph {et~al.}(2026)\citenamefont {Stewart-Wiese}, \citenamefont {Dutta},\ and\ \citenamefont {Hutson}}]{stewartwiese:2026hubbard}%
  \BibitemOpen
  \bibfield  {author} {\bibinfo {author} {\bibfnamefont {C.}~\bibnamefont {Stewart-Wiese}}, \bibinfo {author} {\bibfnamefont {J.}~\bibnamefont {Dutta}}, \ and\ \bibinfo {author} {\bibfnamefont {J.~M.}\ \bibnamefont {Hutson}},\ }\href {\doibase 10.48550/arXiv.2607.28533} {\enquote {\bibinfo {title} {{Hubbard physics with ultracold polar molecules: on-site interaction energies for shielded molecules}},}\ } (\bibinfo {year} {2026}),\ \Eprint {http://arxiv.org/abs/2607.28533} {arXiv:2607.28533 [cond-mat.quant-gas]} \BibitemShut {NoStop}%
\bibitem [{\citenamefont {Stevenson}\ \emph {et~al.}(2024)\citenamefont {Stevenson}, \citenamefont {Singh}, \citenamefont {Elkamshishy}, \citenamefont {Bigagli}, \citenamefont {Yuan}, \citenamefont {Zhang}, \citenamefont {Greene},\ and\ \citenamefont {Will}}]{stevenson:2024}%
  \BibitemOpen
  \bibfield  {author} {\bibinfo {author} {\bibfnamefont {I.}~\bibnamefont {Stevenson}}, \bibinfo {author} {\bibfnamefont {S.}~\bibnamefont {Singh}}, \bibinfo {author} {\bibfnamefont {A.}~\bibnamefont {Elkamshishy}}, \bibinfo {author} {\bibfnamefont {N.}~\bibnamefont {Bigagli}}, \bibinfo {author} {\bibfnamefont {W.}~\bibnamefont {Yuan}}, \bibinfo {author} {\bibfnamefont {S.}~\bibnamefont {Zhang}}, \bibinfo {author} {\bibfnamefont {C.~H.}\ \bibnamefont {Greene}}, \ and\ \bibinfo {author} {\bibfnamefont {S.}~\bibnamefont {Will}},\ }\href {\doibase 10.1103/PhysRevLett.133.263402} {\bibfield  {journal} {\bibinfo  {journal} {Phys. Rev. Lett.}\ }\textbf {\bibinfo {volume} {133}},\ \bibinfo {pages} {263402} (\bibinfo {year} {2024})}\BibitemShut {NoStop}%
\bibitem [{\citenamefont {Chen}\ \emph {et~al.}(2024)\citenamefont {Chen}, \citenamefont {Biswas}, \citenamefont {Eppelt}, \citenamefont {Schindewolf}, \citenamefont {Deng}, \citenamefont {Shi}, \citenamefont {Yi}, \citenamefont {Hilker}, \citenamefont {Bloch},\ and\ \citenamefont {Luo}}]{xinyu+shi}%
  \BibitemOpen
  \bibfield  {author} {\bibinfo {author} {\bibfnamefont {X.-Y.}\ \bibnamefont {Chen}}, \bibinfo {author} {\bibfnamefont {S.}~\bibnamefont {Biswas}}, \bibinfo {author} {\bibfnamefont {S.}~\bibnamefont {Eppelt}}, \bibinfo {author} {\bibfnamefont {A.}~\bibnamefont {Schindewolf}}, \bibinfo {author} {\bibfnamefont {F.}~\bibnamefont {Deng}}, \bibinfo {author} {\bibfnamefont {T.}~\bibnamefont {Shi}}, \bibinfo {author} {\bibfnamefont {S.}~\bibnamefont {Yi}}, \bibinfo {author} {\bibfnamefont {T.~A.}\ \bibnamefont {Hilker}}, \bibinfo {author} {\bibfnamefont {I.}~\bibnamefont {Bloch}}, \ and\ \bibinfo {author} {\bibfnamefont {X.-Y.}\ \bibnamefont {Luo}},\ }\href {\doibase 10.1038/s41586-023-06986-6} {\bibfield  {journal} {\bibinfo  {journal} {Nature}\ }\textbf {\bibinfo {volume} {626}},\ \bibinfo {pages} {283} (\bibinfo {year} {2024})}\BibitemShut {NoStop}%
\bibitem [{\citenamefont {Singh}\ and\ \citenamefont {Greene}(2026)}]{singh:2026}%
  \BibitemOpen
  \bibfield  {author} {\bibinfo {author} {\bibfnamefont {S.}~\bibnamefont {Singh}}\ and\ \bibinfo {author} {\bibfnamefont {C.~H.}\ \bibnamefont {Greene}},\ }\href {\doibase 10.1103/qlvv-r93m} {\bibfield  {journal} {\bibinfo  {journal} {Phys. Rev. Lett.}\ }\textbf {\bibinfo {volume} {136}},\ \bibinfo {pages} {233402} (\bibinfo {year} {2026})}\BibitemShut {NoStop}%
\bibitem [{\citenamefont {Prieschl}\ \emph {et~al.}(2026)\citenamefont {Prieschl}, \citenamefont {Schindewolf},\ and\ \citenamefont {Zillich}}]{prieschl:2026}%
  \BibitemOpen
  \bibfield  {author} {\bibinfo {author} {\bibfnamefont {F.}~\bibnamefont {Prieschl}}, \bibinfo {author} {\bibfnamefont {A.}~\bibnamefont {Schindewolf}}, \ and\ \bibinfo {author} {\bibfnamefont {R.~E.}\ \bibnamefont {Zillich}},\ }\href@noop {} {\  (\bibinfo {year} {2026})},\ \Eprint {http://arxiv.org/abs/2609.01940} {arXiv:2609.01940 [cond-mat.quant-gas]} \BibitemShut {NoStop}%
\bibitem [{\citenamefont {B{\"u}chler}\ \emph {et~al.}(2007)\citenamefont {B{\"u}chler}, \citenamefont {Micheli},\ and\ \citenamefont {Zoller}}]{buechler:2007}%
  \BibitemOpen
  \bibfield  {author} {\bibinfo {author} {\bibfnamefont {H.~P.}\ \bibnamefont {B{\"u}chler}}, \bibinfo {author} {\bibfnamefont {A.}~\bibnamefont {Micheli}}, \ and\ \bibinfo {author} {\bibfnamefont {P.}~\bibnamefont {Zoller}},\ }\href {\doibase 10.1038/nphys678} {\bibfield  {journal} {\bibinfo  {journal} {Nat. Phys.}\ }\textbf {\bibinfo {volume} {3}},\ \bibinfo {pages} {726} (\bibinfo {year} {2007})}\BibitemShut {NoStop}%
\bibitem [{\citenamefont {{\'{S}}mia{\l}kowski}\ and\ \citenamefont {Tomza}(2021)}]{smialkowski:2021}%
  \BibitemOpen
  \bibfield  {author} {\bibinfo {author} {\bibfnamefont {M.}~\bibnamefont {{\'{S}}mia{\l}kowski}}\ and\ \bibinfo {author} {\bibfnamefont {M.}~\bibnamefont {Tomza}},\ }\href {\doibase 10.1103/PhysRevA.103.022802} {\bibfield  {journal} {\bibinfo  {journal} {Phys. Rev. A}\ }\textbf {\bibinfo {volume} {103}},\ \bibinfo {pages} {022802} (\bibinfo {year} {2021})}\BibitemShut {NoStop}%
\bibitem [{\citenamefont {Bonnes}\ \emph {et~al.}(2010)\citenamefont {Bonnes}, \citenamefont {B{\"u}chler},\ and\ \citenamefont {Wessel}}]{bonnes:2010}%
  \BibitemOpen
  \bibfield  {author} {\bibinfo {author} {\bibfnamefont {L.}~\bibnamefont {Bonnes}}, \bibinfo {author} {\bibfnamefont {H.}~\bibnamefont {B{\"u}chler}}, \ and\ \bibinfo {author} {\bibfnamefont {S.}~\bibnamefont {Wessel}},\ }\href {\doibase 10.1088/1367-2630/12/5/053027} {\bibfield  {journal} {\bibinfo  {journal} {New J. Phys.}\ }\textbf {\bibinfo {volume} {12}},\ \bibinfo {pages} {053027} (\bibinfo {year} {2010})}\BibitemShut {NoStop}%
\bibitem [{\citenamefont {Bonnes}\ and\ \citenamefont {Wessel}(2011)}]{bonnes:2011b}%
  \BibitemOpen
  \bibfield  {author} {\bibinfo {author} {\bibfnamefont {L.}~\bibnamefont {Bonnes}}\ and\ \bibinfo {author} {\bibfnamefont {S.}~\bibnamefont {Wessel}},\ }\href {\doibase 10.1103/PhysRevB.83.134511} {\bibfield  {journal} {\bibinfo  {journal} {Phys. Rev. B}\ }\textbf {\bibinfo {volume} {83}},\ \bibinfo {pages} {134511} (\bibinfo {year} {2011})}\BibitemShut {NoStop}%
\bibitem [{\citenamefont {Mazza}\ \emph {et~al.}(2010)\citenamefont {Mazza}, \citenamefont {Rizzi}, \citenamefont {Lewenstein},\ and\ \citenamefont {Cirac}}]{mazza:2010}%
  \BibitemOpen
  \bibfield  {author} {\bibinfo {author} {\bibfnamefont {L.}~\bibnamefont {Mazza}}, \bibinfo {author} {\bibfnamefont {M.}~\bibnamefont {Rizzi}}, \bibinfo {author} {\bibfnamefont {M.}~\bibnamefont {Lewenstein}}, \ and\ \bibinfo {author} {\bibfnamefont {J.~I.}\ \bibnamefont {Cirac}},\ }\href {\doibase 10.1103/PhysRevA.82.043629} {\bibfield  {journal} {\bibinfo  {journal} {Phys. Rev. A}\ }\textbf {\bibinfo {volume} {82}},\ \bibinfo {pages} {043629} (\bibinfo {year} {2010})}\BibitemShut {NoStop}%
\bibitem [{\citenamefont {Singh}\ and\ \citenamefont {Mishra}(2016)}]{singh:2016}%
  \BibitemOpen
  \bibfield  {author} {\bibinfo {author} {\bibfnamefont {M.}~\bibnamefont {Singh}}\ and\ \bibinfo {author} {\bibfnamefont {T.}~\bibnamefont {Mishra}},\ }\href {\doibase 10.1103/PhysRevA.94.063610} {\bibfield  {journal} {\bibinfo  {journal} {Phys. Rev. A}\ }\textbf {\bibinfo {volume} {94}},\ \bibinfo {pages} {063610} (\bibinfo {year} {2016})}\BibitemShut {NoStop}%
\bibitem [{\citenamefont {Axilrod}\ and\ \citenamefont {Teller}(1943)}]{axilrod:1943}%
  \BibitemOpen
  \bibfield  {author} {\bibinfo {author} {\bibfnamefont {B.~M.}\ \bibnamefont {Axilrod}}\ and\ \bibinfo {author} {\bibfnamefont {E.}~\bibnamefont {Teller}},\ }\href {\doibase 10.1063/1.1723844} {\bibfield  {journal} {\bibinfo  {journal} {J. Chem. Phys.}\ }\textbf {\bibinfo {volume} {11}},\ \bibinfo {pages} {299} (\bibinfo {year} {1943})}\BibitemShut {NoStop}%
\bibitem [{\citenamefont {Muto}(1943)}]{muto:1943}%
  \BibitemOpen
  \bibfield  {author} {\bibinfo {author} {\bibfnamefont {Y.}~\bibnamefont {Muto}},\ }\href@noop {} {\bibfield  {journal} {\bibinfo  {journal} {Proc. Phys.-Math. Soc. Jpn.}\ }\textbf {\bibinfo {volume} {17}},\ \bibinfo {pages} {629} (\bibinfo {year} {1943})}\BibitemShut {NoStop}%
\bibitem [{\citenamefont {Micheli}\ \emph {et~al.}(2007)\citenamefont {Micheli}, \citenamefont {Pupillo}, \citenamefont {B{\"u}chler},\ and\ \citenamefont {Zoller}}]{micheli:2007}%
  \BibitemOpen
  \bibfield  {author} {\bibinfo {author} {\bibfnamefont {A.}~\bibnamefont {Micheli}}, \bibinfo {author} {\bibfnamefont {G.}~\bibnamefont {Pupillo}}, \bibinfo {author} {\bibfnamefont {H.~P.}\ \bibnamefont {B{\"u}chler}}, \ and\ \bibinfo {author} {\bibfnamefont {P.}~\bibnamefont {Zoller}},\ }\href {\doibase 10.1103/PhysRevA.76.043604} {\bibfield  {journal} {\bibinfo  {journal} {Phys. Rev. A}\ }\textbf {\bibinfo {volume} {76}},\ \bibinfo {pages} {043604} (\bibinfo {year} {2007})}\BibitemShut {NoStop}%
\bibitem [{\citenamefont {Yang}\ and\ \citenamefont {Karman}(2026)}]{gitlab}%
  \BibitemOpen
  \bibfield  {author} {\bibinfo {author} {\bibfnamefont {H.}~\bibnamefont {Yang}}\ and\ \bibinfo {author} {\bibfnamefont {T.}~\bibnamefont {Karman}},\ }\href {https://gitlab.science.ru.nl/tkarman/three-body-interactions-with-microwave-shielded-molecules} {\enquote {\bibinfo {title} {Three-body interactions with microwave-shielded molecules},}\ } (\bibinfo {year} {2026}),\ \bibinfo {note} {python code for evaluation of two-body and three-body effective potentials for ultracold molecules with double microwave shielding.}\BibitemShut {Stop}%
\bibitem [{\citenamefont {Dutta}\ \emph {et~al.}(2025)\citenamefont {Dutta}, \citenamefont {Mukherjee},\ and\ \citenamefont {Hutson}}]{Dutta_PRR}%
  \BibitemOpen
  \bibfield  {author} {\bibinfo {author} {\bibfnamefont {J.}~\bibnamefont {Dutta}}, \bibinfo {author} {\bibfnamefont {B.}~\bibnamefont {Mukherjee}}, \ and\ \bibinfo {author} {\bibfnamefont {J.~M.}\ \bibnamefont {Hutson}},\ }\href {\doibase 10.1103/PhysRevResearch.7.023164} {\bibfield  {journal} {\bibinfo  {journal} {Phys. Rev. Res.}\ }\textbf {\bibinfo {volume} {7}},\ \bibinfo {pages} {023164} (\bibinfo {year} {2025})}\BibitemShut {NoStop}%
\bibitem [{\citenamefont {J\'o\ifmmode~\acute{z}\else \'{z}\fi{}wiak}\ \emph {et~al.}(2026)\citenamefont {J\'o\ifmmode~\acute{z}\else \'{z}\fi{}wiak}, \citenamefont {Stevenson}, \citenamefont {Will},\ and\ \citenamefont {Karman}}]{jozwiakmap2026}%
  \BibitemOpen
  \bibfield  {author} {\bibinfo {author} {\bibfnamefont {H.~J.}\ \bibnamefont {J\'o\ifmmode~\acute{z}\else \'{z}\fi{}wiak}}, \bibinfo {author} {\bibfnamefont {I.}~\bibnamefont {Stevenson}}, \bibinfo {author} {\bibfnamefont {S.}~\bibnamefont {Will}}, \ and\ \bibinfo {author} {\bibfnamefont {T.}~\bibnamefont {Karman}},\ }\href {\doibase 10.1103/jmnn-glrn} {\bibfield  {journal} {\bibinfo  {journal} {Phys. Rev. A}\ }\textbf {\bibinfo {volume} {114}},\ \bibinfo {pages} {033315} (\bibinfo {year} {2026})}\BibitemShut {NoStop}%
\bibitem [{\citenamefont {Yang}\ \emph {et~al.}()\citenamefont {Yang}, \citenamefont {Dizer}, \citenamefont {Schmidt}, \citenamefont {Christianen},\ and\ \citenamefont {Karman}}]{yang:2026polaron}%
  \BibitemOpen
  \bibfield  {author} {\bibinfo {author} {\bibfnamefont {H.}~\bibnamefont {Yang}}, \bibinfo {author} {\bibfnamefont {E.}~\bibnamefont {Dizer}}, \bibinfo {author} {\bibfnamefont {R.}~\bibnamefont {Schmidt}}, \bibinfo {author} {\bibfnamefont {A.}~\bibnamefont {Christianen}}, \ and\ \bibinfo {author} {\bibfnamefont {T.}~\bibnamefont {Karman}},\ }\href@noop {} {\enquote {\bibinfo {title} {Polarons in microwave-shielded vibrational mixtures of ultracold molecules},}\ }\bibinfo {note} {In preparation}\BibitemShut {NoStop}%
\end{thebibliography}%

\clearpage

\begin{widetext}
\appendix
\section{Effective Two- and Three-Body Potentials}

Microwave-dressed polar molecules support both effective pairwise interactions and genuine nonadditive three-body interactions. 
To derive the corresponding effective potentials, we consider three identical molecules labeled $i$, $j$ and $k$, interacting with two microwave fields of polarization $\sigma^+$ and $\pi$. The total three-molecule Hamiltonian is~\cite{karman:2018,karman:2025}
\begin{equation}
\hat H
=
\sum_{x=i,j,k}
\left[
-\frac{\hbar^2}{2m}\nabla_x^2+\hat H^{(x)}
\right]
+\hat H_{{\rm ac},\sigma}
+\hat H_{{\rm ac},\pi}
+\hat V_{ij}
+\hat V_{jk}
+\hat V_{ki},
\tag{S1}
\end{equation}
where $m$ is the molecular mass, $\hat H^{(x)}$ is the monomer Hamiltonian of molecule $x$, $\hat H_{{\rm ac},\nu}$ describes the energy of microwave mode $\nu=\sigma^+,\pi$, and $\hat V_{xy}$ denotes the dipole-dipole interaction between molecules $x$ and $y$, explicitly given below in Eq.~\eqref{s4}.

The monomer Hamiltonian under microwave dressing is
\begin{equation}
\hat H^{(x)}
=
B_{\rm rot}\big[\hat J^{(x)}\big]^2
+\hat H_{{\rm ac},\sigma}^{(x)}
+\hat H_{{\rm ac},\pi}^{(x)},
\tag{S2}
\end{equation}
where $B_{\rm rot}$ is the rotational constant and $\hat J^{(x)}$ is the rotational angular-momentum operator. The molecule-microwave coupling and the corresponding field Hamiltonian are
\begin{equation}
\hat H_{{\rm ac},\nu}^{(x)}
=
-\frac{1}{2}
\frac{E_\nu}{\sqrt{N_{0,\nu}}}
\left[
\hat d_\nu^{(x)}\hat a_\nu
+
\hat d_\nu^{(x)\dagger}\hat a_\nu^\dagger
\right],
\qquad
\hat H_{{\rm ac},\nu}
=
\hbar\omega_\nu
\left(
\hat a_\nu^\dagger\hat a_\nu-N_{0,\nu}
\right),
\tag{S3}
\end{equation}
where $\hat a_\nu^\dagger$ and $\hat a_\nu$ are the photon creation and annihilation operators, the dipole components are $\hat d_\pi = \hat d_z$ and $\hat d_\sigma = -(\hat d_x+i\hat d_y)/\sqrt{2}$, and $N_{0,\nu}$ are the large reference photon numbers.

We determine the microwave-dressed energy levels as eigenstates of the monomer Hamiltonian with the basis set $\lvert J, m_J, N_\sigma, N_\pi \rangle$~\cite{karman:2025}, where $m_J$ is the rotational angular momentum projection. The state of the microwave fields is described in the photon
number basis, $|N_\nu\rangle$, where $N_\nu + N_{0,\nu}$ is the number of photons in field $\nu$. The monomer Hamiltonian can be approximately set up by a minimal basis of three functions
$\{\lvert 0,0,0,0 \rangle$, 
$\lvert 1,1,-1,0 \rangle$, 
$\lvert 1,0,0,-1 \rangle\}$, with the matrix representation
\begin{align}
\mathbf{H}=
\begin{bmatrix}
0 & \dfrac{\hbar}{2}\Omega_{\sigma} & \dfrac{\hbar}{2}\Omega_{\pi} \\
\dfrac{\hbar}{2}\Omega_{\sigma} & -\hbar\Delta_{\sigma} & 0 \\
\dfrac{\hbar}{2}\Omega_{\pi} & 0 & -\hbar\Delta_{\pi}
\end{bmatrix}.
\label{v0}
\tag{S4}
\end{align}
Notice that Eq.~\eqref{v0} is independent of molecular parameters such as $B_{\rm rot}$, thus it has universality of dimensionless ratios $\Omega_\pi/\Omega_\sigma,
\Delta_\sigma/\Omega_{\rm tot},
\Delta_\pi/\Omega_{\rm tot}$. 
If the basis is extended to include the states $|1,-1,1,-2\rangle$ and $|0,0,0,-2\rangle$, a small correction can arise from the weak admixture of $|1,-1,1,-2\rangle$ into the dressed eigenstates via the intermediate state $|0,0,0,-2\rangle$, which connects $|1,-1,1,-2\rangle$ to the minimal three-state basis. Since this intermediate state is separated from the near-resonant manifold by approximately $4B_{\rm rot}$, this effect is not universal, but the resulting correction is typically negligible.

In the asymptotic region, the trimer states are represented as products of the microwave-dressed monomer eigenstates, $|n_i n_j n_k\rangle=|n_i\rangle|n_j\rangle|n_k\rangle$, with energies $E_{n_i n_j n_k}=\varepsilon_{n_i}+\varepsilon_{n_j}+\varepsilon_{n_k}$ and $\varepsilon_{n_x}$ being the energy of $|n_x\rangle$. We assume that all molecules are initially prepared in the upper field-dressed eigenstate $|0\rangle$ of  the monomer Hamiltonian Eq.~\eqref{v0}, so that the incoming trimer state is $|000\rangle$ with $E_{000}=3\varepsilon_0$.

The dipole-dipole interaction between molecules $i$ and $j$ is given by
\begin{equation}
\hat V_{ij}
=
-\frac{\sqrt{30}\,d^2}{R_{ij}^{3}}
\Big[
\big[
C_{1}(\hat r^{(i)})
\otimes
C_{1}(\hat r^{(j)})
\big]^{(2)}
\otimes
C_{2}(\hat{R}_{ij})
\Big]^{(0)}_{0},
\tag{S5}
\label{s4}
\end{equation}
where $d$ is the body-fixed molecular dipole moment, $C_l(\hat{r})$ is the rank-$l$ tensor whose components are the Racah normalized spherical harmonics depending on the polar angles of the vector $\hat r$. 
Microwave shielding produces a repulsive barrier at relatively large intermolecular separations, suppressing access to shorter distances~\cite{karman:2018,karman:2025}. 
At these larger separations, the dipole-dipole coupling is weak compared with the energy splittings between microwave-dressed channels, allowing a perturbative treatment that applies at essentially all distances relevant to shielded ultracold molecular collisions~\cite{deng:2023}. 
This approximation is also well justified in optical lattices, where interaction energies are typically on the kHz scale, much smaller than the splittings on MHz scale between microwave-dressed states.

In this context, we emphasize the difference in parameter regime considered in Ref.~\cite{prieschl:2026}.
In that work, single-field microwave shielding was considered,
and as a result the first-order dipolar interaction cannot be compensated.
At large distances, this first-order interaction will always dominate due to the slower $R^{-3}$ decay,
resulting in a pairwise additive interaction.
Substantial non-additive many-body interactions were found at substantially shorter distances where perturbation theory does not apply,
and which should not affect molecular scattering~\cite{deng:2023}.
In the present work, we consider double microwave shielding with fully or partially compensated dipolar interaction,
so that the nonadditive three-body interaction occurs at the same order as the two-body interaction and is appreciable at all separations.

Treating the dipole-dipole interaction perturbatively, the first-order perturbation only generates pairwise interaction, which is $U_{ij}^{(1)}=\langle00|\hat V_{ij}|00\rangle$ between molecules $i$ and $j$ and proportional to $P_2(\cos\theta_{ij})/R_{ij}^{3}$, where $P_l(z)$ is a Legendre polynomial. As seen by expanding the second-order expression in Eq.~\eqref{2nd}, products involving the same molecular pair, such as $\hat V_{ij}\hat V_{ij}$, give the second-order pairwise interaction, whereas products involving two different pairs, such as $\hat V_{ij}\hat V_{jk}$, give the cross-pair contribution responsible for the three-body interaction discussed below. The same-pair contribution for molecules $i$ and $j$ is $U_{ij}^{(2)}=\sum_{|n_in_j\rangle\neq |00\rangle}|\langle n_i n_j|\hat V_{ij}|00\rangle|^2/(E_{00}-E_{n_in_j})$, where $|n_in_j\rangle$ denotes intermediate two-molecule states other than the incoming state $|00\rangle$. The effective pairwise potential through second order can therefore be written as
\begin{equation}
U_{ij} 
=
A\frac{P_2(\cos\theta_{ij})}{R_{ij}^{3}}
+
\frac{
B+C P_2(\cos\theta_{ij})+D P_4(\cos\theta_{ij})
}{
R_{ij}^{6}
},
\tag{S6}
\label{s6}
\end{equation}
where $\theta_{ij}$ is the angle between the intermolecular axis and the microwave quantization axis, $A$ is obtained by projecting the first-order interaction onto $P_2$, while $B$, $C$, and $D$ are obtained by projecting the second-order contribution onto $P_0$, $P_2$, and $P_4$, respectively. At the compensation point, $A=0$, leaving the $R^{-6}$ pairwise interaction.

We now consider the cross terms such as $\hat V_{ij}\hat V_{jk}$ appearing in Eq.~\eqref{cross} and how they give rise to nonadditive three-body interactions. Since $\hat V_{ij}$ acts only on molecules $i$ and $j$, whereas $\hat V_{jk}$ acts only on $j$ and $k$, the intermediate state has the form $|0n_j0\rangle$. Thus, only the ``shared'' molecule $j$ is virtually excited or de-excited, while molecules $i$ and $k$ remain in the initial dressed state $|0\rangle$. The resulting energy shift depends simultaneously on $\hat R_{ij}$ and $\hat R_{jk}$, and therefore cannot be decomposed into a sum of pairwise interactions.

The microscopic origin of this second-order contribution can be understood from the dipole matrix elements involved in the cross-pair process. Since molecules $i$ and $k$ remain in the initial dressed state, their contributions involve diagonal dipole matrix elements $\langle 0|\hat{\mathbf d}|0\rangle$, whereas molecule $j$ undergoes a virtual transition $|0\rangle\rightarrow|n_j\rangle\rightarrow|0\rangle$, involving the transition dipole matrix elements $\langle 0|\hat{\mathbf d}|n_j\rangle$ and $\langle n_j|\hat{\mathbf d}|0\rangle$. Microwave dressing mixes rotational states of opposite parity, allowing dipole matrix elements between these components to contribute to the diagonal dipole moments of the dressed states and thereby enabling the nonadditive three-body interaction already at second order.

For bare rotational states of definite parity, however, the diagonal matrix elements vanish because the electric dipole operator is odd under parity. Consequently, the second-order cross-pair contribution is forbidden, as molecules $i$ and $k$ cannot return to their initial states after each has been acted upon by a single dipole operator. The leading nonadditive contribution then arises at third order as an Axilrod--Teller--Muto (ATM) type interaction~\cite{axilrod:1943,muto:1943}, associated with the cyclic product $\hat V_{ij}\hat V_{jk}\hat V_{ki}$. In this process, each molecule is acted upon by two dipole operators, allowing it to return to its initial state. The resulting interaction scales as $R_{ij}^{-3}R_{jk}^{-3}R_{ki}^{-3}$, corresponding to $R^{-9}$ for comparable intermolecular separations, in contrast to the $R^{-6}$ scaling of the microwave-induced second-order contribution.

Mixing of opposite parity states occurs not only for dressing with microwave fields, but also static electric fields.
Therefore interactions of the type discused here should also arise for static electric field shielding \cite{gonzalez,matsuda2020resonant},
but they are particularly important for double microwave shielding~\cite{karman:2025} which partially compensates the first-order dipolar interaction that is otherwise dominant and pair-wide additive.

Owing to the rank-2 tensor structure of the dipole-dipole interaction, the geometry dependence can be grouped into three angular components with $|m|=0,1,2$, corresponding to $C_{20}(\hat R)$, $C_{2,\pm1}(\hat R)$ and $C_{2,\pm2}(\hat R)$, respectively. Therefore Eq.~\eqref{s4} can be rewritten as $\hat V_{ij} = \hat V^{(0)}_{ij}+\hat V^{(1)}_{ij}+\hat V^{(2)}_{ij}$, where
\begin{align}
\hat V^{(0)}_{ij} &= -\frac{\sqrt{6}d^2}{R^3_{ij}} \big\{ \langle 1,1,1,-1|20\rangle C_{11}(\hat r^{(i)}) C_{1,-1}(\hat r^{(j)}) C_{20}(\hat{R}_{ij})+ \langle 1010|20\rangle C_{10}(\hat r^{(i)}) C_{10}(\hat r^{(j)}) C_{20}(\hat{R}_{ij}) \nonumber\\
&~~~~~~~~~~~~~~~~~~~~~~~~+ \langle 1,-1,1,1|20\rangle C_{1,-1}(\hat r^{(i)}) C_{11}(\hat r^{(j)}) C_{20}(\hat{R}_{ij}) \big\}\nonumber\\
&=\frac{\hat X_{ij}^{(0)}}{R_{ij}^{3}}
C_{20}(\hat R_{ij}),
\tag{S7}
\end{align}
\begin{align}
\hat V^{(1)}_{ij}& = \frac{\sqrt{6}d^2}{R^3_{ij}} \big\{\langle 1110|21\rangle C_{11}(\hat r^{(i)}) C_{10}(\hat r^{(j)}) C_{2,-1}(\hat{R}_{ij})+ \langle 1011|21\rangle C_{10}(\hat r^{(i)}) C_{11}(\hat r^{(j)}) C_{2,-1}(\hat{R}_{ij}) +\nonumber\\
&~~ \langle 1,-1,1,0|2,-1\rangle C_{1,-1}(\hat r^{(i)}) C_{10}(\hat r^{(j)}) C_{21}(\hat{R}_{ij}) + \langle 1,0,1,-1|2,-1\rangle C_{10}(\hat r^{(i)}) C_{1,-1}(\hat r^{(j)}) C_{21}(\hat{R}_{ij}) \big\}\nonumber\\
&=  \frac{1}{R^3_{ij}}\big [ \hat X_{ij}^{(1)}C_{21}(\hat R_{ij}) + \hat X_{ij}^{(-1)}C_{2,-1}(\hat R_{ij})\big ],
\tag{S8}
\end{align}
\begin{align}
\hat V^{(2)}_{ij}& = -\frac{\sqrt{6}d^2}{R^3_{ij}} \big\{ \langle 1111|22\rangle C_{11}(\hat r^{(i)}) C_{11}(\hat r^{(j)}) C_{2,-2}(\hat{R}_{ij})+ \langle 1,-1,1,-1|2,-2\rangle C_{1,-1}(\hat r^{(i)}) C_{1,-1}(\hat r^{(j)}) C_{22}(\hat{R}_{ij}) \big\}\nonumber\\
& = \frac{1}{R^3_{ij}}\big [ \hat X_{ij}^{(2)}C_{22}(\hat R_{ij}) + \hat X_{ij}^{(-2)}C_{2,-2}(\hat R_{ij})\big ].
\tag{S9}
\end{align}
Here $\hat X_{ij}^{(q)}$ is a two-molecule operator constructed from products of the dipole operators of molecules $i$ and $j$, acting on their internal degrees of freedom. The intermolecular distance and spatial angular dependence are factored out into $R_{ij}^{-3}$ and $C_{2q}(\hat R_{ij})$, respectively.

For the second-order pairwise interaction, the three angular components contribute independently as a consequence of conservation of the total angular momentum projection. With the two microwave fields, the conserved projection for a molecular pair is
$\mathcal{M}=m_{J_i}+m_{J_j}+m_L+N_\sigma$,
where $m_L$ is the projection of the relative orbital angular momentum. Defining the internal projection as
$M_{\rm int}=m_{J_i}+m_{J_j}+N_\sigma$,
the transition to an intermediate dressed state $|n_i n_j\rangle$ that differs from the incoming state $|00\rangle$ by $\Delta M_{\rm int}$ must be accompanied by
$\Delta m_L=-\Delta M_{\rm int}$.
Since $\hat V_{ij}^{(m)}$ contains the spatial tensor components $C_{2,\pm m}(\hat R_{ij})$ and is diagonal in the photon basis set, only the term with
$m=|\Delta M_{\rm int}|$ of $\hat V_{ij}^{(m)}$ contributes to $|\langle n_i n_j|\hat V_{ij}|00\rangle|^2$. In the sum over intermediate states $|n_i n_j\rangle$, each term therefore belongs to a specific $m$ sector, and cross terms between different angular components are absent in the second-order pairwise interaction.

The same selection rule applies to the second-order three-body cross terms. For the process in Eq.~\eqref{cross}, $\hat V_{ij}$ does not act on molecule $k$ and $\hat V_{jk}$ does not act on molecule $i$, so that
\begin{equation}
\langle000|\hat V_{ij}|0n_j0\rangle
=
\langle00|\hat V_{ij}|0n_j\rangle,
\qquad
\langle0n_j0|\hat V_{jk}|000\rangle
=
\langle n_j0|\hat V_{jk}|00\rangle,
\tag{S10}
\label{matr}
\end{equation}
and the product of matrix elements can be reduced to
$\langle00|\hat V_{ij}|0n_j\rangle \langle n_j0|\hat V_{jk}|00\rangle$ (or $\langle00|\hat V_{jk}|n_j0\rangle \langle 0n_j|\hat V_{ij}|00\rangle$).
If the intermediate dressed state $|0n_j\rangle$ differs from $|00\rangle$ by an internal angular momentum projection $\Delta M_j$, conservation of $\mathcal{M}$ restricts both matrix elements in Eq.~\eqref{matr} to the same sector $m=|\Delta M_j|$. Within this sector, one interaction drives the shared molecule from $|0\rangle$ to $|n_j\rangle$, whereas the other returns it to $|0\rangle$, hence the two transitions involve tensor components $C_{2m}$ and $C_{2,-m}$. Considering the two contributions of $\hat V_{ij} \hat V_{jk}$ and $\hat V_{jk}\hat V_{ij}$ in Eq.~\eqref{cross}, we obtain the combination of spherical harmonics products as $C_{2m}(\hat R_{ij})C_{2,-m}(\hat R_{jk})+C_{2m}(\hat R_{jk})C_{2,-m}(\hat R_{ij})=2\mathrm{Re}\big[C_{2m}(\hat{R}_{ij})C_{2,-m}(\hat{R}_{jk})\big]$, which corresponds to the geometrical factors shown in Eq.~\eqref{three}.

Therefore, after factoring out the dependence on the spatial configuration, each $\hat V^{(m)}$ defines a corresponding three-body coefficient $w_m$ in Eq.~\eqref{three}, which is determined solely by the microwave parameters and evaluated within the two-molecule dressed-state basis:
\begin{equation}
w_m
=
2\sum_{q=\pm m}
\sum_{|n\rangle\neq|0\rangle}
\frac{
\big |
\langle 0n|\hat X^{(q)}|00\rangle
\big |^2
}{
E_{00}-E_{0n}
},
\qquad m=0,1,2 .
\tag{S11}
\label{s10}
\end{equation}
Since the molecules are identical, the required two-molecule matrix elements of $\hat X_{ij}^{(q)}$ are independent of which pair of identical molecules is chosen ($ij$, $jk$ or $ki$), we therefore omit the pair indices on $\hat X^{(q)}$ in Eq.~\eqref{s10}.
We emphasize that in our notation the states $|0n\rangle$ are not adapted to exchange symmetry.
The sum in Eq.~\eqref{s10} runs over the intermediate dressed states $|n\rangle\neq|0\rangle$ of the ``shared'' molecule only, and the factor of $2$ accounts for the contributions of $\hat V_{ij}\hat V_{jk}$ and $\hat V_{jk}\hat V_{ij}$, rather than exchange symmetrization.

\section{Numerical Evaluation of the Effective-Potential Coefficients}

We provide Python code for evaluating the effective two- and three-body interactions derived above, including their microwave-dependent coefficients and spatial dependence \cite{gitlab}.
The functions \texttt{twobody\_coefficient} and \texttt{threebody\_coefficient} calculate the pairwise interaction coefficients $A,B,C,D$ in Eq.~\eqref{s6} and the three-body interaction coefficients $w_0,w_1,w_2$ in Eq.~\eqref{s10}, respectively. The inputs of both functions contain the molecular rotational constant $B_{\rm rot}$, the body-fixed dipole moment $d$, the microwave Rabi frequencies $\Omega_\sigma,\Omega_\pi$ and detunings $\Delta_\sigma,\Delta_\pi$. All physical input and output quantities are in atomic units.

The calculations are performed without static electric or magnetic fields, and hyperfine structure is not included. By default, the molecular basis includes rotational states $J=0,1$, and the photon basis includes states with photon numbers ranging from $-4$ to $+2$ relative to the reference photon number $N_{0,\nu}$ for both the $\sigma^+$ and $\pi$ fields. These basis-set truncations can be adjusted through the parameters \texttt{jmax} and \texttt{nphoton} defined at the beginning of the Python code. For each set of molecular and microwave parameters, the code constructs and diagonalizes the microwave-dressed Hamiltonian and evaluates the perturbative matrix elements entering the effective-potential coefficients.

Several auxiliary functions are provided for the numerical calculations. The functions \texttt{s\_eig} and \texttt{sf\_cg} evaluate the eigenvalues and eigenvectors of Hermitian matrices and the Clebsch--Gordan coefficients, respectively. The function \texttt{g\_legen} provides Gauss--Legendre quadrature points and weights for angular integration, while \texttt{sf\_clm1} evaluates Racah-normalized spherical harmonics $C_{lm}$ at $\phi=0$ using the Condon--Shortley phase convention, with the azimuthal dependence included through $e^{im\phi}$.

As discussed above, the dependence on microwave dressing and molecular geometry can be evaluated separately. The function \texttt{threebody\_geometry} takes the Cartesian coordinates of three molecules, with the microwave quantization axis along $\hat z$, and returns the geometrical coefficients $g_0,g_1$ and $g_2$. For a square lattice with lattice constant $a$, the dimensionless geometrical factors in the main text are given by $\gamma_m=a^6g_m$. The function \texttt{threebody\_potential} takes $w_0,w_1,w_2$ and the same Cartesian coordinates as inputs and returns the total nonadditive three-body potential $W=g_0w_0+g_1w_1+g_2w_2$. The corresponding pairwise potential can be evaluated by combining the coefficients $A,B,C,D$ with the Legendre-polynomial factors in Eq.~(\ref{s6}). Together, these functions enable the evaluation of effective two- and three-body potentials for arbitrary microwave parameters and molecular configurations.

\end{widetext}
\end{document}